\documentclass[11pt]{article}
\usepackage{jheppub}
\usepackage{graphicx}
\usepackage{amsmath,amssymb,amstext,amsfonts} 
\usepackage{mathrsfs}
\usepackage[hang]{subfigure}
\usepackage[utf8]{inputenc}
\usepackage[a4paper,top=1.89in, bottom=0in, left=2in, right=0in]{geometry}
\usepackage{xcolor}
\usepackage{braket}
\usepackage{array}
\usepackage[bottom]{footmisc}
\usepackage{tikz}
\usepackage{tabularx}
\usepackage{appendix}
\usepackage{setspace}
\usepackage{enumitem}
\usepackage{float}
\usepackage[numbers]{natbib}

\newcommand{\nn}{\nonumber}

\usepackage{color}

\newcommand{\be}{\begin{equation}}
\newcommand{\ee}{\end{equation}}
\newcommand{\ba}{\begin{eqnarray}}
\newcommand{\ea}{\end{eqnarray}}

\numberwithin{equation}{section}

\begin{document}
\title{Slowly Rotating Black Holes in Einsteinian Quartic Gravities}

\author[1,2]{Gareth Arturo Marks}
\emailAdd{gamarks@uwaterloo.ca}
\author[2]{Robert B. Mann}
\emailAdd{rbmann@uwaterloo.ca}
\author[2,3]{and Damian Sheppard}
\emailAdd{damian\_sheppard@sfu.ca}

\affiliation[1]{
Department of Applied Mathematics and Theoretical Physics,\\ University of Cambridge,
Wilberforce Road, Cambridge CB3 0WA, United Kingdom\\}
\affiliation[2]{Department of Physics and Astronomy, University of Waterloo,\\
 Waterloo, Ontario, N2L 3G1, Canada\\ }
\affiliation[3]{
Department of Physics, Simon Fraser University \\ Burnaby, British Columbia, V5A 1S6, Canada \\}

\abstract{We study slowly rotating black hole solutions in the six independent theories of Einstein Quartic Gravity (EQG) in four dimensions. Unlike in the static case for which all six theories yield the same solution, for rotating black holes we obtain distinct results for five out of the six theories. Working to leading order in the rotation parameter, we find that the equations characterizing these black holes can be reduced to second order for each theory, similar to what has already been done for Einstein Cubic Gravity. We construct approximate and numerical solutions to these equations, and study how physical properties of the solutions such as the angular velocity, photon sphere, black hole shadow, and innermost stable circular orbit are modified, working to leading order in the coupling constant.  }
\maketitle

\section{Introduction}

Many attempts to quantize gravity lead to the inclusion of higher-curvature corrections to the Einstein-Hilbert action. An understanding of the effects of these higher-curvature corrections on different types of solutions is therefore of significant interest. However, it is generally difficult to construct black hole solutions to higher-curvature modifications to general relativity, which often involve complicated sets of fourth-order differential equations. There are exceptions to this, though. Lovelock gravity \cite{lovelock1970,lovelock1971} is such a well-studied class of theories that are ghost-free on any background. However, curvature terms of order $k$ are topological invariants in $D = 2k$ dimensions, and vanish identically for $D < 2k$. Quasi-topological gravity
\cite{oliva2010c,myers2010c}
, a generalization of this class, presents more examples of higher-curvature theories, yet unfortunately is also trivial in four dimensions.

Recently, a new class of theories have been discovered that are neither topological nor trivial in four dimensions and that have the same graviton spectrum as general relativity on constant curvature backgrounds. 
These properties make this class of theories,
  known as \textit{Generalized Quasi-topological Gravity} (GQTG), 
of considerable interest phenomenologically. 
The first representative of this class of theories to be found was Einsteinian Cubic Gravity (ECG)  \cite{ Bueno:2016xff,hennigar2017b,hennigar2017e,Bueno:2016lrh},
which introduces to the action a unique set of terms cubic in the curvature tensor.  
This theory was discovered to admit non-hairy single-function generalizations to the Schwarzchild black hole \cite{hennigar2017e,bueno2017}. A rather thorough analysis of the basic phenomena of both ECG was performed not long ago
\cite{Hennigar:2018hza,Poshteh:2018wqy}, and more recently 
was carried out for 
Generalized Quasi-topological gravities of order four, or 
Einstein Quartic Gravity (EQG)
\cite{Khodabakhshi:2020hny,Khodabakhshi:2020ddv}, which introduces terms quartic in the curvature tensor. Both studies were in the context of spherically symmetric black holes.

The inclusion of rotation involves significant additional complications. In fact, no analytic generalization of the Kerr solution has yet been constructed so far even for the case of Lovelock gravity. 
However, slowly rotating solutions can often be extracted. This is done using a metric of the form
\be
ds^2 = -f(r)dt^2 + \frac{dr^2}{f(r)} +2ar^2p(r)\sin^2(\theta)dtd\phi +r^2[d\theta^2+\sin^2(\theta)d\phi^2]\label{metric}
\ee
and working to linear order in the rotation parameter $a$.
The Kerr solution in Einstein gravity is well-approximated  by this form for slow rotations, and   an analysis of slowly rotating black hole solutions has already been performed for ECG gravity \cite{adair2020}.

In the cubic case, the GQTG Lagrangian density is  unique. By contrast, in the Quartic case, there are four linearly independent such Lagrangian densities with   analogous properties \cite{ahmed2017a}. In addition, in four dimensions (which we specialize to in this paper) an additional two Lagrangian densities appear, so that it is sensible to speak of six seperate theories of Einstein Quartic Gravity (EQG). For static, spherically symmetric, single-function solutions
the same field equation was found to hold for all six theories \cite{ahmed2017a,Sajadi:2022pcz};
hence such black holes cannot be used to phenomenologically distinguish between them. The predictions for  both cubic and quartic GQTGs that  have been
worked out \cite{Hennigar:2018hza,Khodabakhshi:2020hny,Khodabakhshi:2020ddv} thus leave open  the general question   as to how to empirically distinguish the six  theories from one another as well as from the cubic theory. This is particularly pertinent in the context of  gravitational wave-astronomy  as a possible means of comparing astrophysical rotating black holes in Einstein gravity to those predicted by theories containing various higher-curvature corrections.

Motivated by the above, in this paper we study solutions of the form \eqref{metric} in the six linearly independent theories of EQG that appear in four dimensions. We find that the third and fourth theories predict equivalent solutions, but, unlike in the case of static, spherically symmetric black holes, we obtain distinct results for each of the other theories. In Section 2, we show how to reduce the field equations to a set of second-order differential equations, one of which has already been studied in the context of ECG. We give near-horizon and asymptotic approximate solutions to these equations, as well as construct numerical solutions. We also obtain a continued fraction approximations that approximate the numerical solutions very well everywhere outside the event horizon. In Section 3, we study how various physical properties of the solution are modified in each theory, working perturbatively in the coupling constants. In particular, we study the angular momentum of the event horizon and properties of causal geodesics including the photon sphere, photon ring, and innermost stable circular orbit, as well as the modification to the black hole shadow. 

\section{Slowly Rotating Solutions}\label{cavity}

The action for a general EQG in four dimensions is given by 
\be
S=\frac{1}{16\pi}\int d^4x\sqrt{-g}\left[R-\sum_{i=1}^{6}\hat{\lambda}_{(i)}\mathcal{S}_{(4)}^{(i)}\right]
\ee
where $R$ is the Ricci scalar and the $\mathcal{S}_{(4)}^{(i)}$ are the generalized quasi-topological Lagrangian densities whose forms are presented in 
\cite{ahmed2017a}.  We write the ansatz for the slowly rotating black hole in the following form by letting $x=\cos\theta$:
\be
ds^2 = -N(r)^2f(r)dt^2 + \frac{dr^2}{f(r)} +2ar^2p(r)(1-x^2)dtd\phi +r^2\left[\frac{dx^2}{1-x^2}+(1-x^2)d\phi^2\right]
\ee
Note that this form includes three independent functions of the radial coordinate. In Einstein gravity $p(r) = f(r) -1$ and $N(r)=1$
\cite{Gray:2021roq}; we will find that the latter condition holds without loss of generality but  that $p(r)$ and $f(r)$ are two independent functions.

Our goal is to solve the field equations for this ansatz, working to linear order in $a$. We consider each of the six theories separately. The field equation for the $i$th theory is given by

\be
P_{acde}^{(i)}R_b^{cde}-\frac{1}{2}g_{ab}\mathcal{S}_{(4)}^{(i)}-2\nabla^c\nabla^d P_{acdb}^{(i)} = 0
\ee
where
\be
P_{abcd}^{(i)} = \frac{\partial \mathcal{S}_{(4)}^{(i)}}{\partial R^{abcd}} \, .
\ee
The exact expressions for the tensors $P_{acde}^{(i)}$ are lengthy and of little interest by themselves. At linear order in $a$, 
the field equations $\mathcal{E}_{tt}$ and $\mathcal{E}_{rr}$ are equivalent to one another and to the static case; these admit the solution $N(r)=$ constant, and so we may set $N(r)=1$ without loss of generality. 

Unlike the static case, the    $\mathcal{E}_{t\phi}$ equation is no longer identically
zero, but rather  is a complicated fourth-order differential equation in $p(r)$. However, for all six theories, the following combination of the field equations yields a differential equation that is third order in $p(r)$:
\be\label{p-eq}
\frac{r^4}{f(r)}\left[\mathcal{E}_t^\phi-\frac{arp(r)}{2}\frac{d\mathcal{E}_r^r}{dr} \right] = 0
\ee

Furthermore, the resulting equations depend only on the first three derivatives of $p(r)$ and not on the function itself, so that by making the substitution $g(r) = p'(r)$ we end up with a differential equation for each theory that is second-order in $g(r)$.
As in the spherically symmetric case, the field equation determining $f(r)$ is equivalent for each of the six theories. We may write it, setting $G=1$, as \cite{ Bueno:2016xff, hennigar2017e,Bueno:2016lrh}
\begin{align}
2M&=r \left( 1-f(r)\right)  +\frac {24K}{5\,{r}^{3}} \left[  f'(r) r \left( \frac{rf'(r)}{2}+1-f(r)  \right) f(r) f''(r) +3\, f'(r) ^{4}{r}^{2}
\right. \nonumber\\ 
&\left.+8\,r \left( 1+\frac{f(r)}{2}  \right)  f'(r) ^{3}+24\,f(r)  \left( 1-f(r)\right)  f'(r) ^{2} \right]
\end{align}
This is valid for a general EQG theory— $K$ is a linear combination of the six coupling constants, given by
\be
K = \frac{1}{2}\lambda_{(1)}+\frac{5}{4}\lambda_{(2)}+\frac{1}{2}\lambda_{(3)}+\lambda_{(4)}+2\lambda_{(5)}-\lambda_{(6)}
\ee
By contrast, the equations for $g(r)$ are different for each theory— with the exception of those corresponding to $\mathcal{S}_{(4)}^{(3)}$ and $\mathcal{S}_{(4)}^{(4)}$, which are equivalent.  For example, \eqref{p-eq}  for the $\mathcal{S}_{(4)}^{(6)}$ theory yields
\begin{align}
C&= {r}^{4}g(r)+\lambda_{(6)}\left[ -\frac {104\,g(r) }{15\,{r}^{2}} \left( \frac {11\,f(r) {r}^{3}{}f'''(r) }{26}\right) \left( {\frac { f'(r) r}{2}}+1-f(r)  \right) +\frac {{r}^{4} f''(r) ^{2}}{4} \left( \frac {22\,f(r) }{13}+1 \right)  \right. \nonumber\\ 
&\left.+\frac {34\,{r}^{2}f''(r) }{13} \left( {\frac {29\,{r}^{2} f'(r) ^{2}}{136}}+{\frac {5\, f'(r) r}{68} \left( 1-{\frac {144\,f(r) }{5}} \right) }+ \left( 1-{\frac {93\,f(r) }{34}} \right)  \left( 1-f(r)  \right)  \right) 
\right. \nonumber\\ 
&\left.-\frac {32\,{r}^{3} f'(r) ^{3}}{13}-{\frac {40\,{r}^{2} f'(r) ^{2}}{13} \left( 1-{\frac {373\,f(r) }{80}} \right) }-\frac {9\,r \left( 1-f(r)  \right) f'(r) }{13} \left( 1-{\frac {104\,f(r) }{3}} \right) 
\right. \nonumber\\ 
&\left.+\frac {106\,f(r) }{13}-\frac {251 \left( f(r)  \right) ^{2}}{13}+{\frac {132\, \left( f(r)  \right) ^{3}}{13}} \right) +\frac {88\,g'(r) }{15} \left( \frac {20\,rf(r) f''(r)}{11} \left( {\frac {29\, f'(r) r}{40}}+1-f(r)  \right) 
\right. \nonumber\\ 
&\left.\left.+ f'(r)  \left( {\frac {29\,{r}^{2} f'(r) ^{2}}{44}}+{\frac {20\, f'(r) r}{11} \left( 1-{\frac {51\,f(r) }{40}} \right) }+ \left( 1-f(r)  \right)  \left( 1-{\frac {13\,f(r) }{11}} \right)  \right)  \right) 
\right. \nonumber\\ 
&\left.+{\frac {88\,f(r) g''(r) }{15} \left( {\frac {29\, f'(r) r}{22}}+1-f(r)  \right)  \left( {\frac { f'(r) r}{2}}+1-f(r)  \right) } \right] 
\end{align}
where $C$ is a constant of integration. The expressions for the remaining theories are in Appendix A.

In order to agree with the Kerr solution in the weak coupling limit we will find that we must have $C = 6M$; the same turns out to be true in the other  five theories as well.
In the case of small coupling, we may approximately solve the above equations by assuming a perturbative expansion in the coupling constant. This results in the following expansions to second order:
\begin{align}f(r) &=1-{\frac {2M}{r}}+K \left( {\frac {864{M}^{3}}{5{r}^{9}}}-{\frac {1552{M}^{4}}{5{r}^{10}}} \right) +K^{2} \left( -{\frac {69181952{M}^{7}}{25{r}^{19}}}+{\frac {68746752{M}^{6}}{25{r}^{18}}}-{\frac {17044992{M}^{5}}{25{r}^{17}}} \right) 
\\
{r}^{2}p \left( r \right) &=-{\frac {2M}{r}}+\lambda_{(6)}\left({\frac {1728{M}^{3}}{11{r}^{9}}}-{\frac {1552{M}^{4}}{5{r}^{10}}}\right)+\lambda_{(6)}^2\left(-{\frac {58558464{M}^{5}}{95{r}^{17}}}+{\frac {65788416{M}^{6}}{25{r}^{18}}}-{\frac {69181952{M}^{7}}{25{r}^{19}}}\right) 
\end{align}

\subsection{Asymptotic solution}
We begin with an asymptotic solution in the large-r region. This can be done by taking power series ansatz for $f(r)$ and $g(r)$ :

\begin{align}\label{fperturb}
f_{1/r}(r)&= 1- \frac{2M}{r}+\sum_{n=0}\frac{a_n}{r^n}
\\g_{1/r}(r)&= + \frac{6M}{r^4}+\sum_{n=0}\frac{b_n}{r^n}
\end{align}
The coefficients $a_n$ and $b_n$ determine the behaviour of the solution in the large $r$ region. There is also an homogeneous part to the solution, discussed for $f(r)$ in \cite{Khodabakhshi:2020hny}, but it decays super-exponentially and can therefore be neglected for our purposes. Inserting these ansatz into the field equations, for $f(r)$ we find
\be
f_{1/r}(r)= 1- \frac{2M}{r}+\frac{864}{5}\frac{KM^3}{r^9}-\frac{1552}{5}\frac{KM^4}{r^{10}} + O\left(r^{-17}\right)
\ee
For $g(r)$, the six theories produce different asymptotic expansions, though all have their first correction appear at order $r^{12}$ :
\begin{align}
g_{1/r}^{(1)}(r)&=  \frac{6M}{r^4}+1728\frac{\lambda_{(1)}M^3}{r^{12}}-\frac{9312}{5}\frac{\lambda_{(1)}M^4}{r^{13}}+O\left(r^{-20}\right)
\\
g_{1/r}^{(2)}(r)&=  \frac{6M}{r^4}+2592\frac{\lambda_{(2)}M^3}{r^{12}}-4656\frac{\lambda_{(2)}M^4}{r^{13}}+O\left(r^{-20}\right)
\\
g_{1/r}^{(3)}(r)&=  \frac{6M}{r^4}+1728\frac{\lambda_{(3)}M^3}{r^{12}}-\frac{9312}{5}\frac{\lambda_{(3)}M^4}{r^{13}}+O\left(r^{-20}\right)
\\
g_{1/r}^{(4)}(r)&=  \frac{6M}{r^4}+3456\frac{\lambda_{(4)}M^3}{r^{12}}-\frac{18624}{5}\frac{\lambda_{(4)}M^4}{r^{13}}+O\left(r^{-20}\right)
\\
g_{1/r}^{(5)}(r)&=  \frac{6M}{r^4}+3456\frac{\lambda_{(5)}M^3}{r^{12}}-\frac{37428}{5}\frac{\lambda_{(5)}M^4}{r^{13}}+O\left(r^{-20}\right)
\\
g_{1/r}^{(6)}(r)&=  \frac{6M}{r^4}+1728\frac{\lambda_{(6)}M^3}{r^{12}}-\frac{18624}{5}\frac{\lambda_{(6)}M^4}{r^{13}}+O\left(r^{-20}\right)
\end{align}
where the  $\mathcal{S}_{(4)}^{(3)}$ and $\mathcal{S}_{(4)}^{(4)}$ 
are equivalent upon making the substitution $\lambda_{(3)} = 2\lambda_{(4)}$. Some of the remaining inequivalent theories coincidentally   give equivalent asymptotic expansions, as is the case with  $\mathcal{S}_{(4)}^{(1)}$ and $\mathcal{S}_{(4)}^{(3)}$.

Before moving on, let us say a few words regarding the homogeneous part of the solution for $g(r)$ in the asymptotic region. To do this we insert $g^{(i)}(r) = g_{1/r}^{(i)}(r) + g_h^{(i)}(r)$ into the field equation 
and keep only the terms that are most significant at large $r$, with $g_h^{(i)}(r)$ the homogeneous part of the solution. The resulting equation for $g_h^{(i)}(r)$ takes the following form for all six theories:
\be
M^2\alpha_{(i)}\lambda_{(i)}g_h''(r)-2\frac{M^2\alpha_{(i)}\lambda_{(i)}g_h'(r)}{r}+r^6g_h(r)=0
\ee
where there is no summation over the index $(i)$.
Here each $\alpha_{(i)}$ is a positive constant whose value depends on the theory in question. The general solution to this equation is
\be
g_h^{(i)}(r) = r^{3/2}\left[\tilde{A}I_{3/8}\left(\frac{r^4}{4M\sqrt{\alpha_{(i)}\lambda_{(i)}}}\right)+\tilde{B}K_{3/8}\left(\frac{r^4}{4M\sqrt{\alpha_{(i)}\lambda_{(i)}}}\right) \right]
\ee
where $I_\nu(x)$ and $K_\nu(x)$ are the modified Bessel functions of the first and second kinds, respectively, and $\tilde{A}$ and $\tilde{B}$ are constants of integration. Absorbing various constants into the new constants into $A$ and $B$, we may approximate this to first order at large $r$ as the sum of a super-exponentially growing and decaying mode:
\be
g_h^{(i)}(r)= \left[A\exp\left(\frac{r^4}{4M\sqrt{\alpha_{(i)}\lambda_{(i)}}}\right) + B\exp\left(-\frac{r^4}{4M\sqrt{\alpha_{(i)}\lambda_{(i)}}}\right)\right]\left(\sqrt{\frac{1}{r}} + O\left(r^{-9/2}\right)\right)
\ee
Asymptotic flatness demands that we set $A = 0$, while the super-exponentially decaying mode clearly falls off far more rapidly than our particular solution in powers of $1/r$, and can therefore be neglected.

As a final note, when we integrate each of the $g_{1/r}^{(6)}(r)$ to obtain the asymptotic form of $p(r)$ in each theory a constant of integration will appear, which we may call $\frac{\Omega_\infty}{a}$. This  is related to
the asymptotic angular velocity of spacetime, as 
\be
\Omega = \frac{g_{t \phi}}{g_{\phi \phi}} = -ap(r) \rightarrow \Omega_\infty
\ee
By suitable choice of Killing coordinates $t$ and $\phi$ we may always set $\Omega_\infty = 0$ and from now on we will make that choice. This amounts to starting the integration for $p(r)$ at $r = \infty$.

\subsection{Near horizon solution}

We now wish to solve the equations of motion near the event horizon of the black hole. We begin using an ansatz for $f(r)$ given by
\be
f_{nh}(r) = 4\pi T(r-r_+) +\sum_{n=2}^\infty a_n(r-r_+)^n
\ee
chosen to ensure $f$ has a zero of first order at the horizon $r=r_+$, with $T = f'(r_+)/4\pi$ the usual Hawking temperature. Substituting this into the field equation we obtain the following two equations determining $T$ and $M$ from the linear and quadratic-order parts:
\be
2M={\frac {768K{\pi}^{4}{T}^{4}}{5r_+}}+{\frac {512K{\pi}^{3}{T}^{3}}{5{r_+}^{2}}}+r_+
\ee

\be
{\frac {256\lambda{\pi}^{4}K{T}^{4}}{5{ r_+}^{2}}}+{\frac {512{\pi}^{3}K{T}^{3}}{5{{ r_+}}^{3}}}-4\pi Tr_+ +1=0
\ee
These equations can be solved for $M$ and $T$ in terms of $r_+$; the resulting expressions are large but are presented in \cite{Khodabakhshi:2020hny}. Higher-order terms in the field equation do not determine the parameter $a_2$ in the above. However, the remaining coefficients $a_n$ for $n > 2$ can be determined by unwieldy expressions involving $K$, $T$, $M$, and $a_2$.

We now use a similar power ansatz for $g(r)$, with the only significant difference being that  $g(r)$  is not required to vanish at the event horizon. Using the ansatz 

\be
g_{nh}(r) = \sum_{n=0}^\infty g_n(r-r_+)^n
\ee
we obtain (rather large) equations giving $g_n$ for $n > 0$ in terms of $K$, $T$, $M$, $g_0$, and near horizon coefficients up to $a_{n+1}$ at most. Like $a_2$, $g_0$ is an apparently free parameter. The values of both will be numerically determined in the following section, on the basis of producing a smooth fit with the asymptotic approximation already discussed.

\subsection{Numerical solution and continued fraction approximation}

Neither the asymptotic nor near horizon solutions are valid for all $r$, and so it is desirable to have a means of interpolating between them. Let us begin by constructing a numerical solution for $f(r)$; the procedure for $g(r)$ will be essentially the same. The idea is to use the initial data from the near horizon expansion in a numerical solver that works both outwards and inwards from the horizon radius. Specifically, for the initial data we write
\begin{align}
f(r_+ + \epsilon)&= 4\pi T \epsilon+a_2\epsilon^2
\nonumber\\
f'(r_+ + \epsilon)&= 4\pi T +2a_2\epsilon
\end{align}
 where $\epsilon$ is a small quantity, chosen to be positive for the outer solution and negative for the inner one. A generic choice of $a_2$ will unphysically excite the super-exponentially growing mode. To avoid this and get the asymptotically flat solution, $a_2$ must be chosen with high precision. There is generally a unique value of $a_2$ at which the numerically generated solution agrees with the asymptotic solution to a high degree of accuracy. To find this $a_2$, we must first choose a value $r_\text{max}$ with the requirement that the asymptotic solution is a good approximation for  $r > r_\text{max}$. Then $f(r)$ is recursively solved numerically using different values of $a_2$ until one is found for which $f(r)$ agrees with the numerical solution at $r_\text{max}$ to  the desired precision. Inevitably even this solution will diverge super-exponentially at some point, though that point can be pushed  out to larger values of $r$  by increasingly more precise choices of $a_2$. A full solution valid for all $r$ is obtained by switching to the asymptotic approximation at $r_\text{max}$. 

\begin{figure}[h]
\centering
	\includegraphics[width=0.55\textwidth]{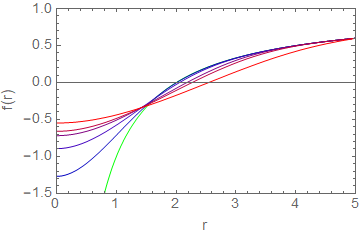}
	\caption{$f(r)$ against $r$ (in units of $M$) for varying values of the coupling constant. The green curve is Einstein gravity, while from blue to red we have $K = 0.1, 1, 5, 10, 50$ 
	  }.
	\label{fig:Fig1}	
\end{figure}

We plot the numerical solution for $f(r)$  in Figure \ref{fig:Fig1} for different values of the coupling $K$.  We note that it has two interesting properties: first, that $f(r)$ remains bounded rather than diverging to $-\infty$ as $r\to 0$, and second, that increasing $K$ increases the event horizon radius $r_+$.

The idea for constructing the numerical solution for $g(r)$ is the same as for $f(r)$. We construct initial data via
\begin{align}
g(r_+ + \epsilon)&= g_0 + g_1\epsilon +g_2\epsilon^2
\nonumber\\
g'(r_+ + \epsilon)&= g_1 +2g_2\epsilon
\end{align}
using our near horizon expansion to rewrite $g_1$ and $g_2$ in terms of $g_0$ and the already-determined $a_2$. Just as for $a_2$, there is a unique choice of $g_0$ that avoids exciting the super-exponentially growing mode. We find it in precisely an analogous manner, and construct a full solution for $g(r)$ by switching from the numerical solution evaluated for this choice of $g_0$ to the asymptotic solution at a large enough $r$ that the latter becomes a good approximation.

\begin{figure}[h]
	\;\;\includegraphics[width=0.49\textwidth]{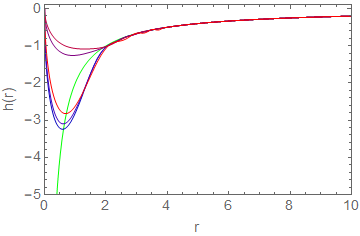}\quad\;\;\;\;\includegraphics[width=0.49\textwidth]{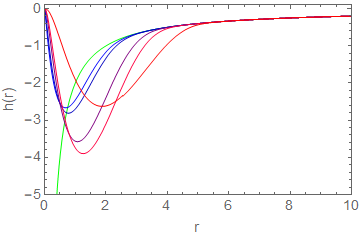}
	\newline
	\includegraphics[width=0.505\textwidth]{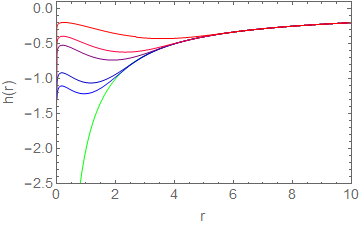}\quad\;\includegraphics[width=0.505\textwidth]{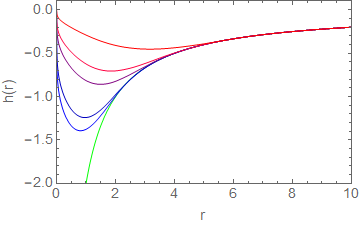}
	\caption {$h(r)$ against $r$ (in units of $M$) for varying values of the coupling constant. From left to right, top to bottom, we show theories $\mathcal{S}_{(4)}^{(2)}$, $\mathcal{S}_{(4)}^{(4)}$ (which is equivalent to $\mathcal{S}_{(4)}^{(3)}$), $\mathcal{S}_{(4)}^{(5)}$, and $\mathcal{S}_{(4)}^{(6)}$ . In each case the green curve represents Einstein gravity, while 
the remaining curves correspond to 	
 $\lambda_{(i)} = 0.5, 1, 5, 10,$ and $50$, increasing from  blue to red. The exception is theory 2, the top left, for which only a narrow range of $\lambda_{(2)}$ produce a reasonable result: in this case $\lambda_{(2)} = 0.5, 0.6, 0.75, 1, 1.25$ is used.
	}
\label{fig:Fig2}
\end{figure}

With the solution for $g(r)$ in hand, we may numerically integrate it to obtain $p(r)$ and hence $h(r) = r^2p(r)$. We start the integration at $r= \infty$ to $p(r) \rightarrow 0$ as $r \rightarrow \infty$. As discussed, this is equivalent to choosing a frame that does not rotate at infinity. We show these numerical results in Figure \ref{fig:Fig2} for each of the theories. A few interesting observations emerge. First, each of the theories except for that corresponding to $\mathcal{S}_{(4)}^{(5)}$ approaches zero as $r \rightarrow 0$. This behaviour is also observed in ECG and differs from Einstein gravity. $h(r)$ in the $\mathcal{S}_{(4)}^{(5)}$ theory is unique in that it approaches $-\infty$ in the limit $r \rightarrow 0$, but far more sharply and with manifestly different behaviour compared to Einstein gravity.

The near horizon and asymptotic solutions are only valid in their respective regions, and the numerical solution is computationally intensive to generate and inevitably breaks down at sufficiently large distance. Fortunately, we may use another approximation that yields an excellent analytic approximation to the numerical solution everywhere outside the horizon. This is the continued fraction approximation.  This approximation has been previously employed in the static case in both 
ECG \cite{Hennigar:2018hza}
and EQG \cite{Khodabakhshi:2020hny}
and recently in the slowly rotating case in  quadratic gravity
\cite{Sajadi:2023smm}. 
Since the continued fraction equation for $f(r)$ 
in EQG 
is the same in the slowly rotating case we study, we can employ it here.  
We first perform the change of coordinates
\be
x = 1 - \frac{r_+}{r}
\ee
so that the spacetime interval outside the horizon is in the range $x \in [0,1)$. We then take an ansatz
\be \label{contfracf}
f(x) = x\left[1 - \varepsilon (1-x) + (b_0-\varepsilon)(1-x)^2 +B(x)(1-x)^3 \right]
\ee
where $B(x)$ is given by the continued fraction
\be
\frac{b_1}{1+\frac{b_2x}{1+\frac{b_3x}{1+...}}}
\ee
Inserting this ansatz in the relevant field equation yields
\be
\varepsilon = \frac{2M}{r_+}-1, \quad b_0 = 0
\ee
Furthermore, expanding near the horizon the coefficients $b_i$ are straightforwardly obtained. Keeping terms up to $b_5$ produces an excellent analytic approximation for $f(r)$, valid for all $r$. In particular, the expressions for the first two coefficients are reasonably simple \cite{Khodabakhshi:2020hny}:

\be
b_1=4\pi r_+ T+\frac{4M}{r_+}-3, \quad b_2 = -\frac{r_+^3a_2+16r_+^2T+6(M-r_+)}{4\pi r_+^2T+4M-3r_+}
\ee
All higher-order terms in the continued fraction can be written in terms of $a_2$, $T$, and $r_+$, though the exact expressions are cumbersome, being presented in Appendix C.

To produce an analogous approximation for $g(r)$, a natural first approximation might be to choose an ansatz of the form of \eqref{contfracf} for $h(r) = r^2p(r)$ as it is a similarly dimensionless function. Unfortunately, such an approach leads inevitably to the problem that the continued fraction coefficients are underdetermined --  equating terms at order $r^0$ involves two continued fraction coefficients, equating terms at order $r^1$ involves these and another continued fraction coefficient, and so on.  We can, however, construct a continued fraction approximation for $g(r)$ using the following ansatz:
\be
g(x) = \frac{1}{r^3}\left[-\gamma(1-x) + (d_0 - \gamma)(1-x)^2 + D(x)(1-x)^A\right]
\ee
where
\be
D(x) =\frac{d_1}{1+\frac{d_2x}{1+\frac{d_3x}{1+...}}}
\ee
We can then numerically integrate this to obtain $p(r)$. While the approximation for $p(r)$ is not analytic, it is nonetheless far less computationally intensive to obtain  than constructing a full numerical solution. Note that the exponent $A$ has been left arbitrary. Continued fraction coefficients can be computed with this ansatz for any integer $A > 2$, but we found that $A = 9$ produced the best agreement with the numerical solution everywhere outside the event horizon. Using this ansatz in an asymptotic expansion, along with the known continued fraction expansion for $f(r)$, yields
\be
\gamma = d_0 = -\frac{6M}{r_+}
\ee
To get the continued fraction coefficients $d_i$, we once again expand about the event horizon and compare to the near horizon coefficients. We find that generally $d_n$ can be expressed in terms of $M$, $r_h$, and the near horizon coefficients for $g(r)$ up to $g_{n-1}$, which themselves depend on $g_0$, $T$, and the coupling constant. We must therefore use the value of $g_0$ numerically determined by the shooting method described above. We find that the first two continued fraction coefficients are given by
\be
d_1 = \frac{6M}{r_+} - g_0r_+^3, \quad d_2 = -\frac{24M+(A+3)r_+d_1+g_1r_+^5}{d_1r_+}
\ee
The remaining coefficients are given by similar, longer expressions presented in Appendix C. As with $f(r)$, we keep the first five coefficients in the continued fraction. A comparison of the continued fraction approximation to the numerical solution is shown in Figure 3.

\begin{figure}[h]
	\includegraphics[width=0.505\textwidth]{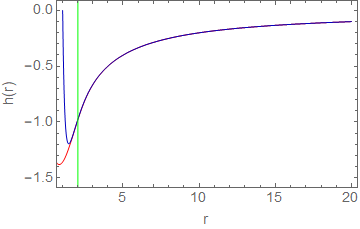}\quad\includegraphics[width=0.505\textwidth]{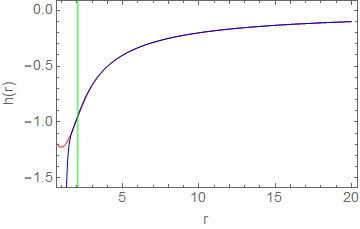}
	\newline
	\includegraphics[width=0.505\textwidth]{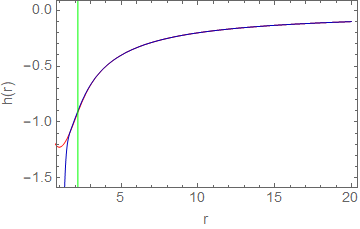}\quad\;\includegraphics[width=0.505\textwidth]{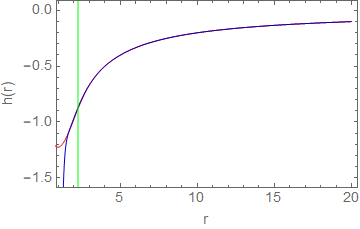}
	\caption {Comparison of the numerical solution (red) to the continued fraction approximation (blue) for $h(r)$ against $r$ (in units of $M$) in the $\mathcal{S}_{(4)}^{(6)}$ theory. From left to right, top to bottom, we take $\lambda_{(1)} = 0.5, 1, 5, 10$. Notice that, unlike the continued fraction for $f(r)$, the continued fraction approximation for $h(r)$ is generally only valid outside the event horizon, indicated on each plot by a green line. The approximation is similarly accurate for the other theories.
	}
\label{fig:Fig3}
\end{figure}

\section{Properties of the solution}

In this section we study various properties of slowly rotating black hole solutions in the six EQG theories. First we determine the angular velocity of the horizon. We then move to the study of geodesics in spacetimes described by these solutions. Much of the theory concerning geodesics using a metric of the form used in this paper is developed in \cite{adair2020}. Throughout, we will present perturbative results for the sixth theory only; similar expressions for the other five theories are given in Appendix D.

\subsection{Angular velocity of the event horizon}
The angular velocity of the event horizon is given by
\be
\Omega = -\frac{g_{t\phi}}{g_{\phi \phi}}\bigg\rvert_{r=r_+}
\ee
With a metric of the form given by \eqref{metric} this becomes
\be
\Omega = -ap(r_+)
\ee
We construct an expression for the angular velocity in the perturbative regime as follows. Setting $f(r_+) = 0$ in \eqref{fperturb} we construct an expression for the horizon radius up to second order in $K$. The result is 
\be
r_+ = 2M - \frac{11}{160}\frac{K}{M^5}-\frac{929}{51200}\frac{K^2}{M^{11}} + O\left({K^3}\right)
\ee
Using the perturbative expression for $p(r)$, then, we obtain the following for each of the six theories:
\begin{align}
\Omega_{(1)} &=\frac{\chi}{M}\left[\frac{1}{4}+\frac{73}{2816}\frac{\lambda_{(1)}}{M^6}+-\frac{14659}{31129600}\frac{\lambda_{(1)}^2}{M^12} \right]
\\
\Omega_{(2)} &=\frac{\chi}{M}\left[\frac{1}{4}-\frac{67}{5632}\frac{\lambda_{(2)}}{M^6}-\frac{162541}{24903680}\frac{\lambda_{(2)}^2}{M^{12}} \right]
\\
\Omega_{(3)} &=\frac{\chi}{M}\left[\frac{1}{4}-\frac{73}{2816}\frac{\lambda_{(3)}}{M^6}+\frac{14659}{31129600}\frac{\lambda_{(3)}^2}{M^{12}} \right]
\\
\Omega_{(4)} &=\frac{\chi}{M}\left[\frac{1}{4}+\frac{73}{1408}\frac{\lambda_{(4)}}{M^6}+\frac{14659}{7782400}\frac{\lambda_{(4)}^2}{M^{12}} \right]
\\
\Omega_{(5)} &=\frac{\chi}{M}\left[\frac{1}{4}-\frac{35}{704}\frac{\lambda_{(5)}}{M^6}-\frac{27677}{1945600}\frac{\lambda_{(5)}^2}{M^{12}} \right]
\\
\Omega_{(6)} &=\frac{\chi}{M}\left[\frac{1}{4}+\frac{35}{1408}\frac{\lambda_{(6)}}{M^6}-\frac{27677}{7782400}\frac{\lambda_{(6)}^2}{M^{12}} \right]
\end{align}
where we have defined $\chi = a/M$, and as before   $\mathcal{S}_{(4)}^{(3)}$ and $\mathcal{S}_{(4)}^{(4)}$ 
are equivalent upon making the substitution $\lambda_{(3)} = 2\lambda_{(4)}$.
These expressions work well in the regime $M \gg \lambda_{(i)}^{1/6}$. For smaller $M$, we must resort to the numerical solutions described in the previous sections. 

\subsection{Geodesics}
We now study the geodesics of the slowly rotating solution.  The theory to linear order in $a$ is developed in \cite{adair2020} for a metric of the form

\be
ds^2 = -f(r)dt^2 + \frac{dr^2}{f(r)} +2ah(r)\sin^2(\theta)dtd\phi +r^2[d\theta^2+\sin^2(\theta)d\phi^2]
\ee
This analysis results in a linear system of equations for the coordinate functions:
\begin{align}
r^2 \dot{t} &= \frac{Er^2 + ah(r)\ell_z}{f(r)}
\\
r^2 \dot{\phi} &= \frac{\ell_z}{\sin ^2\theta} - a\frac{Eh(r)}{f(r)}
\\
r^2 \dot{\theta} &= \pm \sqrt{j^2-\frac{\ell_z}{\sin ^2\theta}}
\\
\dot{r}^2&=-f(r)\left(\xi^2 + \frac{j^2}{r^2}\right) + E^2 +\frac{2ah(r)E\ell_z}{r^2}
\end{align}
Asymptotically, $j^2$ represents the total angular momentum, while $\ell_z$ is the component along the $z$-axis (i.e. $\theta = 0, \pi$). $\xi$ is the norm of the tangent vector, where
\be
\xi^2 = -g_{a b}\dot{x}^a\dot{x}^b
\ee
\subsubsection{The Photon Sphere}
For null geodesics we have $\xi^2 = 0$ and we may always choose $E = 1$. The equation for the radial coordinate can then be written
\be
\dot{r^2}+V_{ph}(r) = 0 \quad \text{where}\quad V_{ph}(r)=\frac{j^2f(r) - 2a\ell_zh(r)}{r^2} - 1
\ee
The photon sphere is formed by constant-$r$ photon orbits, which appear when 
\be
V_{ph}(r_{ps})=V'_{ph}(r_{ps})=0
\ee
To first order in $a$ we expand $r_{ps}$ and $j^2_{ps}$ according to
\be
r_{ps} = r_{ps}^{(0)} + ar_{ps}^{(1)}, \quad j^2_{ps} = (j_{ps}^{(0)})^2 + (aj_{ps}^{(1)})^2
\ee

With this expansion $r_{ps}^{(0)}$ is determined by 
\be \label{rpsEqn}
r_{ps}^{(0)}f'\left(r_{ps}^{(0)}\right) = 2f\left(r_{ps}^{(0)}\right)
\ee
while the other photon sphere parameters can be evaluated in terms of this quantity:
\be
r_{ps}=r_{ps}^{(0)}+ \frac{2a\ell_zf(rh'-2h)}{r(r^2f''-2f)}\bigg\rvert_{r=r_{ps}^{(0)}}
\ee
\be
j_{ps}^2 = \frac{(r_{ps}^{(0)})^2}{f(r_{ps}^{(0)})}+\frac{2a\ell_zh(r_{ps}^{(0)})}{f(r_{ps}^{(0)})}
\ee
In the perturbative regime we may use the solutions \eqref{fperturb} for $f(r)$ and $h(r)$ to evaluate these to second order in the coupling constant. We do this via a method analogous to that used to obtain the angular momentum, by first constructing a perturbative approximation for $r_{ps}^{(0)}$ using \eqref{rpsEqn} and using this in the equations for $r_{ps}$ and $j_{ps}^2$. The results for the  $\mathcal{S}_{(4)}^{(6)}$  theory 
are as follows:
\begin{align}
r_{ps}&= 3M-{\frac {1648\lambda_{(6)}}{32805{M}^{5}}}+{\frac {23839744{\lambda_{(6)}}^{2}}{3228504075{M}^{11}}}+a\left( -{\frac {2l_{z}}{9M}}-{\frac {19808l_{z}\lambda_{(6)}}{885735{M}^{7}}}+{\frac {158446592l_{z}{\lambda_{(6)}}^{2}}{29056536675{M}^{13}}} \right) 
\\
 j_{ps}^2&=27{M}^{2}-{\frac {208\lambda_{(6)}}{729{M}^{4}}}+{\frac {1031168{\lambda_{(6)}}^{2}}{39858075{M}^{10}}}+a\left( -4l_{z}-{\frac {3424l_{z}\lambda_{(6)}}{24057{M}^{6}}}+{\frac {4776791552l_{z}{\lambda_{(6)}}^{2}}{224919117225{M}^{12}}} \right) 
\end{align}
with expressions for the other theories given in Appendix~\ref{AppD}. 
Alternatively, when $\lambda_{(i)}/M^6$ becomes of order 1 or larger we must again resort to the numerical solution for an accurate evaluation of these quantities.

\subsection{Geodesics in the equatorial plane}
Here we consider geodesics confined to the equatorial plane, which amounts to specializing the previous case by setting $\theta = \pi/2$ and $\dot{\theta} = 0$. Note that these constraints combine to enforce $j^2 = \ell_z^2$ so that there will be no need in this section to distinguish between these angular momenta. The $r$ equation  can be interpreted as being analogous to that of a particle moving in a potential,
\be\dot{r}^2 + V_{\text{eff}}(r) = 0 \quad \text{where} \quad V_{\text{eff}}(r) = f(r)\left(\xi^2 + \frac{j^2}{r^2}\right) - \frac{2ah(r)Ej}{r^2}-E^2\ee

\subsubsection{Timelike geodesics: ISCO}
First let us consider the case of circular, timelike geodesics, i.e. with $\xi^2 = 1$ and $\dot{r} = 0$. The conditions for these geodesics to exist are

\be V_{\text{eff}}(r) = V'_{\text{eff}}(r) = 0 \ee
The stability of the circular orbit is deduced from the sign of $V''_{\text{eff}}(r)$, with a positive sign indicating stability and a negative one instability. We determine the location of the innermost stable circular orbits (ISCO) by searching for orbits which are inflection points, i.e. for which $V''_{\text{eff}}(r) = 0$. To find these working to linear order in $a$, we write the following expansions for $r$, $j$ and $E$:
\be
r_{\text{ISCO}} = r_{\text{ISCO}}^{(0)} + ar_{\text{ISCO}}^{(1)}, \quad j_{\text{ISCO}} = j_{\text{ISCO}}^{(0)} + aj_{\text{ISCO}}^{(1)}, \quad E_{\text{ISCO}} = E_{\text{ISCO}}^{(0)} + aE_{\text{ISCO}}^{(1)}
\ee
Substituting these in to the equations $V_{\text{eff}}(r) = 0$, $V'_{\text{eff}}(r) = 0$, $V''_{\text{eff}}(r)=0$ and collecting in powers of $a$ results in a system of six equations that must be solved. These are very large and yield little insight on their own, but   the solution procedure, at least in the perturbative regime, is essentially the same for all six theories. We present a perturbative solution for the $\mathcal{S}_{(4)}^{(6)}$ theory:
\begin{align}
r_{\text{ISCO}} &= 6M-{\frac {2221\lambda_{(6)}}{209952{M}^{5}}}-{\frac {82796797{\lambda_{(6)}}^{2}}{1322395269120{M}^{11}}}
\\ \nonumber &\mp  a\left(-{\frac {4\sqrt {6}}{3}}-{\frac {9715\lambda_{(6)}\sqrt {6}}{629856{M}^{6}}}-{\frac {15486310769{\lambda_{(6)}}^{2}\sqrt {6}}{79343716147200{M}^{12}}}\right)
\\
j_{\text{ISCO}}&= \pm\left(2\sqrt {3}M-\frac{1373\sqrt {3}\lambda_{(6)}}{3149280{M}^{5}}-\frac {38293721\sqrt {3}\lambda_{(6)}^{2}}{19835929036800{M}^{11}}\right)
\\ \nonumber &+ a\left(-{\frac {2\sqrt {2}}{3}}-{\frac {13621\sqrt {2}\lambda_{(6)}}{6298560{M}^{6}}}-{\frac {3080166109\sqrt {2}{\lambda_{(6)}}^{2}}{158687432294400{M}^{12}}}\right)
\\
E_{\text{ISCO}}&={\frac {2\sqrt {2}}{3}}-{\frac {191\sqrt {2}\lambda_{(6)}}{6298560{M}^{6}}}-{\frac {26522027\sqrt {2}{\lambda_{(6)}}^{2}}{158687432294400{M}^{12}}}
\\ \nonumber &\mp a\left(-{\frac {\sqrt {3}}{54M}}-{\frac {445363\sqrt {3}\lambda_{(6)}}{3741344640\,{M}^{7}}}-{\frac {91154747183\sqrt {3}{\lambda_{(6)}}^{2}}{74622765036441600{M}^{13}}}
\right)
\end{align}
Here taking the positive sign corresponds to prograde orbits ($j_\text{ISCO}^{(0)} > 0$) while the negative sign corresponds to retrograde orbits (($j_\text{ISCO}^{(0)} < 0$). Once again,  when $\lambda_{(i)}/M^6$ becomes of order 1 or larger we must resort to the numerical solutions for an accurate reporting of these corrections.

\subsubsection{Null geodesics: photon rings}
We now consider how rotation deforms the photon rings of the black hole. These are constant-r orbits describing null geodesics in the equatorial plane $\theta = \pi/2$. We therefore seek the simultaneous zeroes of the effective potential and its first derivative. Instead of $E$ and $j$, we work with the angular velocity $\omega = d\phi/dt$,  which is conserved along the photon trajectory. The resultant equations determining the location of the photon rings read
\begin{align}
0 &= \omega^2r^2 +2a \omega h(r) - f(r)
\\0 &= 2\omega^2r +2a \omega h'(r) - f'(r)
\end{align}
The solution to these equations to first order in $a$ is given by
\begin{align}
r_{\text{pr}\pm} &= r_{\text{ps}} \pm a \frac{2\sqrt{f( r_{\text{ps}})}[ r_{\text{ps}}h'( r_{\text{ps}})-2h( r_{\text{ps}})]}{2f( r_{\text{ps}})- r_{\text{ps}}^2f''( r_{\text{ps}})}
\\
\omega_{\text{pr}\pm} &= \mp\frac{\sqrt{f(r_{\text{ps}})}}{r_{\text{ps}}} - a\frac{h(r_{\text{ps}})}{r_{\text{ps}}^2}
\end{align}
In these expressions $r_{\text{ps}}$ is the radius of the photon sphere in the static, spherically symmetric solution, obtained by solving the equation
\be
\frac{r_{\text{ps}}f'(r_{\text{ps}})}{2}-f(r_{\text{ps}})=0
\ee
Also, the positive sign again corresponds to the prograde photon ring, and the minus sign to the retrograde photon ring. Similar to what we have done previously, we can construct a perturbative solution for the photon ring parameters by first solving this equation to second order in the coupling constant. The resulting perturbative solution takes the following form for the $\mathcal{S}_{(4)}^{(6)}$ theory:
\begin{align}
r_{\text{pr}\pm} &= 3M-{\frac {1648\lambda_{(6)}}{32805{M}^{5}}}+{\frac {23839744{\lambda_{(6)}}^{2}}{3228504075{M}^{11}}}\\ \nonumber &\pm a\left( {\frac {2\sqrt {3}}{3}}+{\frac {6256\sqrt {3}\lambda_{(6)}}{98415{M}^{6}}}-{\frac {158876608\sqrt {3}{\lambda_{(6)}}^{2}}{9685512225{M}^{12}}} \right) 
\\
\omega_{\text{pr}\pm}&=\pm\left(-{\frac {\sqrt {3}}{9\,M}}-{\frac {104\,\sqrt {3}\lambda_{(6)}}{177147\,{M}^{7}}}+{\frac {1411552\,\sqrt {3}{\lambda_{(6)}}^{2}}{29056536675\,{M}^{13}}}\right)\\ \nonumber & + a\left( {\frac {2}{27\,{M}^{2}}}+{\frac {19984\,\lambda_{(6)}}{5845851\,{M}^{8}}}-{\frac {85092352\,{\lambda_{(6)}}^{2}}{198746710857\,{M}^{14}}} \right) 
\end{align}
with expressions for the other theories given in Appendix~\ref{AppD}.

\subsection{Black Hole Shadow}

We now turn to a discussion of the black hole shadow. Similar to our study of geodesics, the general form of the black hole shadow is derived for theories of this type in \cite{adair2020}. Consider, without loss of generality, an observer at a distance $r_0$ at  polar angle $\theta_0$ and azimuthal angle $\phi_0 = 0$. The observer then receives a photon moving along a trajectory with increasing $r$. Taking $\alpha$ as the angle between the photon trajectory and the azimuthal direction, and $\pi/2 - \delta$ as the angle of incidence of the photon on the plane $r = r_0$,  the contour $\delta(\alpha)$ of the black hole shadow is approximately a 
circle of  radius $R_{\text{sh}}$ centered at $\alpha = 0$, $r_0\delta = D $, where \cite{adair2020}
\be
R_{\text{sh}} = \frac{r_\text{ps}^{(0)}}{\sqrt{f\left(r_\text{ps}^{(0)}\right)}}, \quad D = -\frac{a \sin \theta_0 h\left(r_\text{ps}^{(0)}\right)}{f\left(r_\text{ps}^{(0)}\right)}.
\ee

The effect of the rotation is hence to shift the black hole shadow a distance $D$ from the radial direction.

A perturbative solution for $R$ and $D$ in $\mathcal{S}_{(4)}^{(6)}$  is given by the following:
\begin{align}
R_{\text{sh}}&=3\sqrt {3}M-{\frac {104\sqrt {3}}{6561{M}^{5}}}\lambda_{(6)}+{\frac {4505056\sqrt {3}}{3228504075{M}^{11}}}{\lambda_{(6)}}^{2} 
\\
D &= a\sin(\theta_0)\left(-2-{\frac {1712}{24057{M}^{6}}}\lambda_{(6)}+{\frac {2388395776}{224919117225{M}^{12}}}{\lambda_{(6)}}^{2}\right)
\end{align}
Similar perturbative expressions for the other five theories are given in Appendix~\ref{AppD}.

\section{Conclusion}

We have  constructed  slowly rotating black hole solutions for all six Einstein Quartic Gravity theories $\mathcal{S}_{(4)}^{(J)}$
in 4 spacetime dimensions, with $J=1..6$ In the spherically symmetric case, all six theories yield the same metric, but in the slowly rotating case this degeneracy is broken.  Five of the six theories yield distinct solutions, with $\mathcal{S}_{(4)}^{(3)}$ and $\mathcal{S}_{(4)}^{(4)}$ having degenerate soutions.  

Our results indicate that investigations of photon orbits and shadows of rotating black holes could be used to distinguish EQG theories from ECG, and even distinct EQG theories from each other (apart from the degeneracy noted above).  Of course the general GQT action could be a linear combination of ECG, EQG, and other higher curvature theories, and so actual observations will in practice have to measure (or set bounds on) the various coefficients appearing in the ISCO, ring, and shadow parameters in Appendix~\ref{AppD}.  Observed dependence of these quantities on mass can be used to distinguish the order of the curvature. Obtaining distinctions between parameters having the same mass dependence will be more challenging, and will have to rely on detailed statistical analysis of a broad range of black holes, as well as perhpas consideration of higher-order rotational corrections.

As with the cubic theory
\cite{adair2020}, we find that   the order-reduction phenomenon observed is a general property  in the slowly rotating case  as well as the spherically symmetric case.
This suggests that slowly-rotating solutions in a general GQT theory will exhibit this feature. It would be of interest to consider this problem in general.  Unlike Lovelock gravity, 
in which slowly rotating solutions are 
completely characterized by the metric for the static solution, slowly rotating solutions in EQG (and ECG \cite{adair2020}) do not have this feature. Understanding the conditions in which this is manifest is an interesting problem for further investigation.

 Finally, an investigation of   the thermodynamic properties of  slowly rotating black holes would be of interest. However the entropy and   temperature exhibit dependence on the rotation parameter only at order $a^2$, and so a proper study of thermodynamic behaviour would necessitate computing all relevant quantities to (at least) this order.
 A full solution for  arbitrary values of the rotation parameter would be 
 of greatest interest, but of course of considerably greater difficulty.

 \section*{Acknowledgments}
 This work was supported in part by the Natural Sciences and Engineering Council of Canada. We are grateful to R.A. Hennigar for helpful discussions.

\appendix
\section{Appendix A: Field equations for $g$}
\label{AppA}

Here we present explicit forms of the field equation for $g(r)$ in all theories, with the expression for the $C6$ theory also being shown in the main text.
Note that the equations for theories 3 and 4 are identical.

Theory 1:
\begin{align} 
C &= -\frac {58\,\lambda_{(1)}}{9\,{r}^{2}} \left[  \left( \frac { \left( f'(r)  \right) r}{2}+1-f \left( r \right)  \right)  \left( \frac {37\, \left( f'(r)  \right) r}{290}+1-f \left( r \right)  \right) {r}^{2}f \left( r \right) g''(r) \right.\\ \nonumber &\left.
- \frac {{r}^{3}g \left( r \right) f \left( r \right) f'''(r) }{2} \left( {\frac { \left( f'(r)  \right) r}{2}}+1-f \left( r \right)  \right) -\frac {15\,{r}^{4}g \left( r \right)  \left( f''(r)  \right) ^{2}}{29} \left( \frac {29\,f \left( r \right) }{30}+1 \right) \right.\\ \nonumber & \left.
+\frac {91\,{r}^{2}f''(r) }{145} \left( f \left( r \right) r \left( {\frac {37\, \left( f'(r)  \right) r}{182}}+1-f \left( r \right)  \right) g'(r) +\frac {6\,g \left( r \right) }{91} \left( -\frac {37\,{r}^{2} \left( f'(r)  \right) ^{2}}{24}\right.\right.\right.\\ \nonumber & \left.\left.\left.
-\frac {r \left( 1-1035\,f \left( r \right)  \right) f'(r) }{12}+ \left( 1+118\,f \left( r \right)  \right)  \left( 1-f \left( r \right)  \right)  \right)  \right) + \left( f'(r)  \right)  \left( \frac {37\,{r}^{2} \left( f'(r)  \right) ^{2}}{580}\right.\right.\\ \nonumber & \left.
+\left.\frac {91\, \left( f'(r)  \right) r}{145} \left( 1-\frac {327\,f \left( r \right) }{182} \right) + \left( 1-\frac {344\,f \left( r \right) }{145} \right)  \left( 1-f \left( r \right)  \right)  \right) {r}^{2}g'(r)  \right.\\\ \nonumber &\left.
-\frac {60\,g \left( r \right) }{29} \left( -\frac {19\,{r}^{3} \left( f'(r)  \right) ^{3}}{300}-\frac {817\,{r}^{2} \left( f'(r)  \right) ^{2}}{600} \left( 1-\frac {1982\,f \left( r \right) }{817} \right) -\frac {133\,r \left( 1-f \left( r \right)  \right) f'(r) }{100} 
\left( 1-\frac {853\,f \left( r \right) }{133} \right)  \right.\right.\\ \nonumber &\left.\left. +1+\frac {19\,f \left( r \right) }{5}-\frac {53\, \left( f \left( r \right)  \right) ^{2}}{5}+\frac {29\, \left( f \left( r \right)  \right) ^{3}}{5} \right)    \right]  -{r}^{4}g \left( r \right)
\end{align}

Theory 2:
\begin{align} 
C &=  6\,\frac {\lambda_{(2)}}{{r}^{2}} \left[  \left( -\frac{1}{2}rf'(r) +1-f \left( r \right)  \right)  \left( \frac{1}{2}rf'(r) +1-f \left( r \right)  \right) f \left( r \right) {r}^{2}g''(r) \right.\\ \nonumber& \left.
 -\frac{1}{2}g \left( r \right)  \left( \frac{1}{2}rf'(r) +1-f \left( r \right)  \right) f \left( r \right) {r}^{3}f'''(r) -\frac{2}{3}g \left( r \right)  \left( \frac{3}{4}f \left( r \right) +1 \right) {r}^{4} \left( f''(r)  \right) ^{2}\right.\\ \nonumber & \left.
-\frac{1}{2} \left( f \left( r \right)  \left( f'(r)  \right)  \left( g'(r)  \right) {r}^{2}-\frac{2}{3}g \left( r \right)  \left( \frac{3}{4} \left( f'(r)  \right) ^{2}{r}^{2}-\frac{3}{2}r \left( 1-4\,f \left( r \right)  \right) f'(r)\right.\right.\right.\\ \nonumber & \left.\left.\left.
 + \left( 1+\frac{9}{2}\,f \left( r \right)  \right)  \left( 1-f \left( r \right)  \right)  \right)  \right) {r}^{2}f''(r) + \left( -\frac{1}{4} \left( f'(r)  \right) ^{2}{r}^{2}-\frac{1}{2}f \left( r \right)  \left( f'(r)  \right) r+ \right.\right.\\ \nonumber & \left.\left.
\left( 1-f \left( r \right)  \right)  \left( 1-3\,f \left( r \right)  \right)  \right)  \left( f'(r)  \right) {r}^{2}g'(r) -\frac{8}{3}g \left( r \right)  \left( \frac{1}{2}{r}^{3} \left( f'(r)  \right) ^{3} -\frac { \left( 9-\frac{9}{2}f \left( r \right)  \right) {r}^{2} \left( f'(r)  \right) ^{2}}{8}\right.\right.\\ \nonumber & \left.\left.
-\frac {15\,r \left( 1-f \left( r \right)  \right) f'(r) }{8} \left( 1-\frac {12\,f \left( r \right) }{5} \right) +1+5/2\,f \left( r \right) -8\, \left( f \left( r \right)  \right) ^{2}+\frac{9}{2} \left( f \left( r \right)  \right) ^{3} \right)  \right] -g \left( r \right) {r}^{4}
\end{align}

Theories 3,4:
\begin{align} 
C &= -\frac {24\lambda_{(4)}}{5\,r} \left[ \frac {f \left( r \right)}{{r}^{2}}  \left( f'(r)  \right)  \left( rf'(r) -2\,f \left( r \right) +2 \right) g''(r) {2}+r \left( rf \left( r \right)  \left( rf'(r) +1-f \left( r \right)  \right)\right) g'(r)\right.\\ \nonumber & \left.
  -\frac {g \left( r \right) }{2} \left( {r}^{2} \left( f'(r)  \right) ^{2}+\frac {8\,rf'(r) }{3} \left( 1-{\frac {15\,f \left( r \right) }{8}} \right) +\frac {34-34\,f \left( r \right) }{3} \left( 1-{\frac {9\,f \left( r \right) }{17}} \right)  \right)   \,f''(r) \right.\\ \nonumber &\left.
+\frac { \left( {r}^{2} \left( f'(r)  \right) ^{2}+2\,r \left( 1-f \left( r \right)  \right) f'(r) +2\,f \left( r \right)  \left( 1-f \left( r \right)  \right)  \right) r \left( f'(r)  \right) \,g'(r) }{2}\right.\\ \nonumber & \left.
-\frac {5\,g \left( r \right) }{24} \left( -\frac {56\,{r}^{2}\, \left( f'(r)  \right) ^{3}}{5}-\frac {44\,r\, \left( f'(r)  \right) ^{2}}{5} \left( 1-\frac {51\,f \left( r \right) }{11} \right) -\frac {16\,\,  f'(r) }{5} \left( 1-{\frac {27\,f \left( r \right) }{2}} \right) \right)  \right] -r^4g(r)
\end{align}

Theory 5:
\begin{align} 
C &= -\frac {182\,\lambda_{(5)}}{15\,{r}^{2}} \left[ \frac {f \left( r \right)  \left(  \left( f' \left( r \right)  \right) r+2-2\,f \left( r \right)  \right) {r}^{2}g' \left( r \right) }{2} \left(  \left( f' \left( r \right)  \right) r+{\frac{38}{91}}-{\frac {38\,f \left( r \right) }{91}} \right) 
\right.\\ \nonumber& \left.
-\frac {19\,f \left( r \right) g \left( r \right) {r}^{3} \left(  \left( f' \left( r \right)  \right) r+2-2\,f \left( r \right)  \right) f'''(r) }{182}+\frac {4\,{r}^{4}g \left( r \right)  \left( f''(r)  \right) ^{2}}{91} \left( -\frac {19\,f \left( r \right) }{4}+1 \right) 
\right.\\ \nonumber & \left.
+{r}^{2} \left( rf \left( r \right)  \left(  \left( f' \left( r \right)  \right) r+{\frac{110}{91}}-{\frac {110\,f \left( r \right) }{91}} \right) g'(r) -\frac {g \left( r \right) }{2} \left( {r}^{2} \left( f' \left( r \right)  \right) ^{2}-\frac {10\, \left( f' \left( r \right)  \right) r}{91} \left( 1+\frac {351\,f \left( r \right) }{5} \right) 
\right.\right.\right. \\ \nonumber & \left.\left.\left. 
+\frac {304-304\,f \left( r \right) }{91} \left( 1-\frac {111\,f \left( r \right) }{38} \right)  \right)  \right) f''(r) +\frac {{r}^{2} \left( f' \left( r \right)  \right) g'(r) }{2} \left( {r}^{2} \left( f' \left( r \right)  \right) ^{2}+\frac {220\, \left( f' \left( r \right)  \right) r}{91} \left( 1-{\frac {129\,f \left( r \right) }{110}} \right) 
\right.\right.\\ \nonumber & \left.\left.
+\frac {76-76\,f \left( r \right) }{91} \left( 1-{\frac {2\,f \left( r \right) }{19}} \right)  \right) -\frac {15\,g \left( r \right) }{182} \left( -\frac {412\,{r}^{3} \left( f' \left( r \right)  \right) ^{3}}{15}-\frac {58\,{r}^{2} \left( f' \left( r \right)  \right) ^{2}}{3} \left( 1-{\frac {992\,f \left( r \right) }{145}} \right) 
\right. \right. \\ \nonumber & \left. \left.
+\frac { \left( 52-52\,f \left( r \right)  \right) rf' \left( r \right) }{5} \left( 1+{\frac {241\,f \left( r \right) }{13}} \right) -{\frac{32}{15}}+\frac {976\,f \left( r \right) }{15}-\frac {1856\, \left( f \left( r \right)  \right) ^{2}}{15}+ \frac {304\, \left( f \left( r \right)  \right) ^{3}}{5}  \right) \right]-{r}^{4}g \left( r \right)
\end{align}

\section{Appendix B: Coefficients for the Near Horizon Expansion}
Here we present the coefficients $g_2$ and $g_1$ for each of the six theories. $g_0$ must be numerically determined, while the expressions for $g_3$ and up are much larger.

Theory 1:
\begin{align}
g_1 =  &\frac {1}{8\,\pi\,Ta\lambda_1\,{{r_+}}^{2} \left( 148\,{{\it 
rh}}^{2}{\pi}^{2}{T}^{2}+364\,\pi\,T{r_+}+145 \right) } \left[ -45
\,a{{r_+}}^{6}g_{0}+592\,g_{0}\, \left( {\pi}^{2}{T}^{2}+\frac {75
\,a_{2}}{74} \right) a\lambda_1\,a_{2}\,{{r_+}}^{4}\right. \\ \nonumber 
&\left.-2432\,g_{0}\,a
\lambda_1\, \left( {\pi}^{2}{T}^{2}-{\frac {a_{2}}{304}} \right) \pi\,T
{{r_+}}^{3}+ \left( -13072\,g_{0}\, \left( {\pi}^{2}{T}^{2}+{\frac 
{3\,a_{2}}{1634}} \right) \lambda_1\,a+90\,C \right) {{r_+}}^{2}\right. \\ \nonumber 
&\left.-
3192\,\pi\,Tag_{0}\,\lambda_1\,{r_+}+600\,ag_{0}\,\lambda_1 
\right]
\end{align}

\begin{align}
g_2 =  &\frac {1}{4736\,{\pi}^{3}{T}^{3}\lambda_1\,{{r_+}}^{5}+11648
\,{\pi}^{2}{T}^{2}\lambda_1\,{{r_+}}^{4}+4640\,\lambda_1\,\pi\,T{{
r_+}}^{3}} \left[ -45\,{{r_+}}^{7}g_{1}-180\,g_{0}\,{{r_+}}^{
6}\right. \\ \nonumber 
&\left.-2368\,\lambda_1\, \left( {T}^{2} \left( a_{2}\,g_{1}-{\frac {273\,a_
{3}\,g_{0}}{74}} \right) {\pi}^{2}-{\frac {91\,T\pi\,{a_{2}}^{2}g_{0}
}{74}}-{\frac {75\,a_{2}\, \left( a_{2}\,g_{1}+6\,a_{3}\,g_{0}
 \right) }{296}} \right) {{r_+}}^{5}\right. \\ \nonumber 
&\left.+16128\,\lambda_1\, \left( {T}^{
3}{\pi}^{3}g_{1}-{\frac {139\,{T}^{2}{\pi}^{2}a_{2}\,g_{0}}{63}}-{
\frac {545\,\pi\,T}{2016} \left( a_{2}\,g_{1}-{\frac {438\,a_{3}\,g_{0
}}{545}} \right) }+{\frac {43\,{a_{2}}^{2}g_{0}}{576}} \right) {{\it
rh}}^{4}\right. \\ \nonumber 
&\left.+124416\, \left( {T}^{3}{\pi}^{3}g_{0}-{\frac {7\,{T}^{2}{\pi}
^{2}g_{1}}{2592}}-{\frac {3037\,T\pi\,a_{2}\,g_{0}}{15552}}-{\frac {
151\,a_{2}\,g_{1}}{31104}}-{\frac {a_{3}\,g_{0}}{1728}} \right) 
\lambda_1\,{{r_+}}^{3}\right. \\ \nonumber 
&\left.+94656\,\lambda_1\, \left( {T}^{2}{\pi}^{2}g_{0
}-{\frac {133\,T\pi\,g_{1}}{3944}}-{\frac {133\,a_{2}\,g_{0}}{7888}}
 \right) {{r_+}}^{2}+12312\,\lambda_1\, \left( g_{0}\,\pi\,T+{\frac 
{25\,g_{1}}{513}} \right) {r_+}-1200\,g_{0}\,\lambda_1 \right]
\end{align}

Theory 2:
\begin{align}
g_1 =  &\frac {1}{96\,{\pi}^{3}{T}^{3}a\lambda_2\,{{r_+}}^{4}-24\,\pi
\,Ta\lambda_2\,{{r_+}}^{2}} \left[ a{{r_+}}^{6}g_{0}+48\,g_{0}\,
 \left( {\pi}^{2}{T}^{2}-a_{2}/3 \right) aa_{2}\,\lambda_2\,{{r_+}}^
{4} \right. \\ \nonumber 
&\left. -512\,g_{0}\, \left( {\pi}^{2}{T}^{2}+{\frac {3\,a_{2}}{64}}
 \right) a\lambda_2\,T\pi\,{{r_+}}^{3}+ \left( 288\,g_{0}\, \left( {
\pi}^{2}{T}^{2}+{\frac {a_{2}}{72}} \right) \lambda_2\,a-2\,C \right) {
{r_+}}^{2}+120\,\pi\,Tag_{0}\,\lambda_2\,{r_+}\right. \\ \nonumber 
&\left.-16\,ag_{0}\,
\lambda_2 \right] 
\end{align}

\begin{align}
g_2 =  &\frac {1}{384\,{\pi}^{3}{T}^{3}\lambda_2\,{{r_+}}^{5}-96\,\pi
\,T\lambda_2\,{{r_+}}^{3}} \left[ {{r_+}}^{7}g_{1}+4\,g_{0}\,{{
r_+}}^{6}-192\, \left( {\pi}^{2}{T}^{2}g_{1}+1/12\,a_{2}\,g_{1}+1/2
\,a_{3}\,g_{0} \right) \lambda_2\,a_{2}\,{{r_+}}^{5}\right. \\ \nonumber 
&\left.-896\, \left( {T
}^{3}{\pi}^{3}g_{1}+{\frac {9\,{T}^{2}{\pi}^{2}a_{2}\,g_{0}}{28}}+{
\frac {3\,T \left( a_{2}\,g_{1}+6\,a_{3}\,g_{0} \right) \pi}{112}}+{
\frac {11\,{a_{2}}^{2}g_{0}}{224}} \right) \lambda_2\,{{r_+}}^{4}\right. \\ \nonumber 
&\left.-
1088\,\lambda_2\, \left( {T}^{3}{\pi}^{3}g_{0}+{\frac {3\,{\pi}^{2}{T}^
{2}g_{1}}{34}}-{\frac {49\,T\pi\,a_{2}\,g_{0}}{136}}-{\frac {a_{2}\,g_
{1}}{68}}-{\frac {3\,a_{3}\,g_{0}}{272}} \right) {{r_+}}^{3}\right. \\ \nonumber 
&\left.-1632\,
 \left( {T}^{2}{\pi}^{2}g_{0}-{\frac {5\,T\pi\,g_{1}}{68}}-{\frac {5\,
a_{2}\,g_{0}}{136}} \right) \lambda_2\,{{r_+}}^{2}-280\,\lambda_2\,
 \left( \pi\,Tg_{0}+{\frac {2\,g_{1}}{35}} \right) {r_+}+32\,g_{0}
\,\lambda_2 \right]
\end{align}

Theories 3, 4:
\begin{align}
g_1 =  &\frac {1}{768\,\lambda_4\,{\pi}^{2}{{r_+}}^{2}a{T}^{2}}
 \left[ 5\,a{{r_+}}^{5}g_{0}+384\,{\pi}^{2}{T}^{2}aa_{2}\,g_{0}\,
\lambda_4\,{{r_+}}^{3}-3584\,\lambda_4\,\pi\, \left( {\pi}^{2}{T}^{2}
-a_{2}/14 \right) g_{0}\,aT{{r_+}}^{2}\right.\\ \nonumber & \left. 
+ \left( -704\, \left( {\pi}^
{2}{T}^{2}-{\frac {17\,a_{2}}{44}} \right) \lambda_4\,g_{0}\,a-10\,C
 \right) {r_+}-64\,\pi\,Tag_{0}\,\lambda_4 \right]  \left( {r_+}
\,\pi\,T+{\frac{1}{2}} \right)^{-1}
\end{align}

\begin{align}
g_2 =  &\frac {1}{3072\,{\pi}^{2}{T}^{2}\lambda_4\,{{r_+}}^{3}}
 \left[ 5\,{{r_+}}^{6}g_{1}+20\,g_{0}\,{{r_+}}^{5}-1536\,\lambda
4\, \left( T \left( a_{2}\,g_{1}-\frac{3}{4}a_{3}\,g_{0} \right) \pi-\frac{1}{4}{a
_{2}}^{2}g_{0} \right) T\pi\,{{r_+}}^{4}\right.\\ \nonumber &\left.
-3584\,\lambda_4\, \left( {T
}^{3}{\pi}^{3}g_{1}+{\frac {51\,{T}^{2}{\pi}^{2}a_{2}\,g_{0}}{28}}+
\frac {5\,\pi\,T}{56} \left( a_{2}\,g_{1}-\frac {12\,a_{3}\,g_{0}}{5}
 \right) -1/28\,{a_{2}}^{2}g_{0} \right) {{r_+}}^{3}\right.\\ \nonumber &\left.
{+9472\,
\lambda_4\, \left( {T}^{3}{\pi}^{3}g_{0}-{\frac {23\,{T}^{2}{\pi}^{2}g_
{1}}{148}}-{\frac {33\,T\pi\,a_{2}\,g_{0}}{148}}+{\frac {17\,a_{2}\,g_
{1}}{592}}+{\frac {51\,a_{3}\,g_{0}}{592}} \right) {{r_+}}^{2}}\right.\\ \nonumber &\left.
+3712
\,\lambda_4\, \left( {T}^{2}{\pi}^{2}g_{0}-\frac {T\pi\,g_{1}}{58}-{
\frac {a_{2}\,g_{0}}{116}} \right) {r_+}+64\,\pi\,Tg_{0}\,\lambda_4
 \right]  \left( {r_+}\,\pi\,T+{\frac{1}{2}} \right) ^{-1}
\end{align}

Theory 5:
\begin{align}
g_1 =  &\frac {1}{16\,\pi\,Ta\lambda\,{{r_+}}^{2} \left( 364\,{{\it
rh}}^{2}{\pi}^{2}{T}^{2}+220\,\pi\,T{r_+}+19 \right) } \left[ 15\,a
{{r_+}}^{6}g_{0}+2912\,g_{0}\,a\lambda\, \left( {\pi}^{2}{T}^{2}-{
\frac {a_{2}}{91}} \right) a_{2}\,{{r_+}}^{4}\right.\\ \nonumber& \left.
-26368\,g_{0}\,Ta
\lambda\,\pi\, \left( {\pi}^{2}{T}^{2}+{\frac {5\,a_{2}}{1648}}
 \right) {{r_+}}^{3}+ \left( -4640\,g_{0}\,\lambda\, \left( {\pi}^{
2}{T}^{2}-{\frac {19\,a_{2}}{145}} \right) a\right.\right.\\ \nonumber & \left.\left.
-30\,C \right) {{r_+}}^
{2}+624\,\pi\,Tag_{0}\,\lambda\,{r_+}-32\,ag_{0}\,\lambda \right]
\end{align}

\begin{align}
g_2 =  &\frac {1}{64\,\pi\,T\lambda\,{{r_+}}^{3} \left( 364\,{{\it
rh}}^{2}{\pi}^{2}{T}^{2}+220\,\pi\,T{r_+}+19 \right) } \left[ 15\,{
{r_+}}^{7}g_{1}+ 60\,g_{0}\,{{r_+}}^{6}-11648\, \left( {T}^{2}
 \left( a_{2}\,g_{1}-{\frac {165\,a_{3}\,g_{0}}{182}} \right) {\pi}^{2
}\right.\right.\\ \nonumber & \left.\left.
-{\frac {55\,T\pi\,{a_{2}}^{2}g_{0}}{182}}+{\frac {a_{2}\, \left( a_{
2}\,g_{1}+6\,a_{3}\,g_{0} \right) }{364}} \right) \lambda\,{{r_+}}^
{5}-21504\,\lambda\, \left( {T}^{3}{\pi}^{3}g_{1}+\frac {335\,\pi\,T}{1344} \left( a_{2}\,
g_{1}-{\frac {42\,a_{3}\,g_{0}}{335}} \right) \right.\right.\\ \nonumber &\left.\left.
+{\frac {13\,{a_{2}}^{2
}g_{0}}{2688}} \right) {{r_+}}^{4}+100608\, \left( {T}^{3}{\pi}^{3}
g_{0}-{\frac {71\,{T}^{2}{\pi}^{2}g_{1}}{1048}}-{\frac {297\,T\pi\,a_{
2}\,g_{0}}{2096}}+{\frac {19\,a_{2}\,g_{1}}{4192}}+{\frac {19\,a_{3}\,
g_{0}}{1048}} \right) \lambda\,{{r_+}}^{3}\right.\\ \nonumber &\left.
+43776\,\lambda\, \left( 
{T}^{2}{\pi}^{2}g_{0}+{\frac {13\,\pi\,Tg_{1}}{912}}+{\frac {13\,a_{2}
\,g_{0}}{1824}} \right) {{r_+}}^{2}+3280\, \left( \pi\,Tg_{0}-{
\frac {2\,g_{1}}{205}} \right) \lambda\,{r_+}+64\,g_{0}\,\lambda \right] 
\end{align}

Theory 6:
\begin{align}
g_1 =  &\frac {1}{3712\,{\pi}^{3}{T}^{3}\lambda\,{{r_+}}^{4}+2560\,{\pi}^{
2}{T}^{2}\lambda\,{{r_+}}^{3}+352\,\lambda\,\pi\,T{{r_+}}^{2}}
 \left[ -15\,g_{0}\,{{r_+}}^{6}+1856\, \left( {\pi}^{2}{T}^{2}+{
\frac {13\,a_{2}}{232}} \right) \lambda\,a_{2}\,g_{0}\,{{r_+}}^{4} \right.\\ \nonumber & \left.
-
16384\, \left( {\pi}^{2}{T}^{2}-{\frac {5\,a_{2}}{512}} \right) 
\lambda\,\pi\,g_{0}\,T{{r_+}}^{3}+ \left( -5120\, \left( {\pi}^{2}{
T}^{2}-{\frac {17\,a_{2}}{160}} \right) g_{0}\,\lambda+90\,M \right) {
{r_+}}^{2} \right.\\ \nonumber &\left.
 -288\,\pi\,Tg_{0}\,\lambda\,{r_+}+104\,g_{0}\,\lambda
 \right] 
\end{align}

\begin{align}
g_2 =  &\frac {1}{128\,\lambda\,\pi\,T{{r_+}}^{3} \left( 116\,{{\it 
rh}}^{2}{\pi}^{2}{T}^{2}+80\,\pi\,T{r_+}+11 \right) } \left[ -15\,{
{r_+}}^{7}g_{1}-60\,g_{0}\,{{r_+}}^{6}\right.\\& \nonumber \left.
-7424\,\lambda\, \left( {T
}^{2} \left( g_{1}\,a_{2}-{\frac {30\,a_{3}\,g_{0}}{29}} \right) {\pi}
^{2}-{\frac {10\,{a_{2}}^{2}\pi\,Tg_{0}}{29}}-{\frac {13\,a_{2}\,
 \left( g_{1}\,a_{2}+6\,a_{3}\,g_{0} \right) }{928}} \right) {{r_+}
}^{5}\right.\\ \nonumber& \left.
-10752\, \left( {\pi}^{3}{T}^{3}g_{1}+{\frac {307\,{\pi}^{2}{T}^{
2}a_{2}\,g_{0}}{84}}+{\frac {115\,\pi\,T}{336} \left( g_{1}\,a_{2}-{
\frac {48\,a_{3}\,g_{0}}{115}} \right) }-{\frac {3\,{a_{2}}^{2}g_{0}}{
112}} \right) \lambda\,{{r_+}}^{4}\right.\\ \nonumber & \left.
{+79104\, \left( {\pi}^{3}{T}^{3}g
_{0}-{\frac {6\,g_{1}\,{\pi}^{2}{T}^{2}}{103}}-{\frac {409\,\pi\,Ta_{2
}\,g_{0}}{2472}}+{\frac {23\,g_{1}\,a_{2}}{4944}}+{\frac {17\,a_{3}\,g
_{0}}{824}} \right) \lambda\,{{r_+}}^{3}}\right.\\ \nonumber & \left.
+41088\,\lambda\, \left( {
\pi}^{2}{T}^{2}g_{0}-{\frac {3\,\pi\,Tg_{1}}{428}}-{\frac {3\,a_{2}\,g
_{0}}{856}} \right) {{r_+}}^{2}+3680\,\lambda\, \left( \pi\,Tg_{0}+
{\frac {13\,g_{1}}{460}} \right) {r_+}-208\,g_{0}\,\lambda \right] 
\end{align}

\section{Appendix C: Continued Fraction Coefficients}
The coefficients $b_3$ and $b_4$ appearing in the continued fraction approximation for $f(r)$ are as follows:
\begin{align}
b_3= - &\frac{1}{9216} \frac{1}{K (\pi r_+ T +\frac{1}{2})(r_+^2 T \pi+M -\frac{3}{4}r_+ )T^2r_+^2 \pi^2 b_2}\Bigl[ 20\pi T (b_2+3) r_+^8+(-15b_2 -30)r_+^7
\nn\\
&\nonumber+(20 M b_2+30M)r_+^6+21504K \Bigl(b_2^2+\frac{160}{21}b_2+\frac{487}{28} \Bigr) T^4 \pi^4 r_+^5-17664 K T^3 \Bigl(b_2^2 +\frac{136}{23}b_2
\nn\\
&\nonumber+\frac{122}{23}\Bigr)\pi^3 r_+^4+33792 K  T^2 \Bigl(M \Bigl(b_2^2+\frac{196}{33} b_2+\frac{74}{11} \Bigr) T\pi-\frac{15}{22}b_2-\frac{57}{44}-\frac{3}{88}b_2^2 \Bigr) \pi^2 r_+^3
\nn\\
&\nonumber-7680 K  T  \Bigl(M T \Bigl (b_2^2-\frac{12}{5} \Bigr) \pi -\frac{9}{40}(b_2+2)^2 \Bigr)\pi r_+^2
\nn\\
&\nonumber-12288\Bigl(-b_2-\frac{3}{2}\Bigr) M K \Bigl(\Bigl (b_2+\frac{3}{2}\Bigr) MT \pi-\frac{3}{8} b_2-\frac{3}{4}\Bigr) T \pi r_+ 
\nn\\
& \nonumber +3072\Bigl(b_2+\frac{3}{2}\Bigr)^2 M^2 K  T \pi \Bigr] \,,
\end{align}
 
\begin{align}
b_4 &= -\,{\frac {1}{24576  \Bigl( \pi \,r_+T+\frac{1}{2} \Bigr){K} b_{3} \Bigl({r_+}^{2}T\pi +M -\frac{3}{4}\,r_+ \Bigr){T}^2{r_+}^4{\pi}^2 b_{2}}}
\nn\\
& \nonumber \times \Bigl[ 20\,T \Bigl( 6+{b_{{2}}}^{2}+ \Bigl( b_{{3}}+4 \Bigr) b_{{2}} \Bigr) 
\mbox{}\pi \,{r_+}^{10}+ \Bigl( -70-15\,{b_{{2}}}^{2}+ \Bigl( -15\,b_{{3}}-60 \Bigr) b_{{2}} \Bigr) {r_+}^{9}
\nn\\
& \nonumber\mbox{}+20\,M \Bigl( 4+{b_{{2}}}^{2}+ \Bigl( b_{{3}}+4 \Bigr) b_{{2}} \Bigr) 
\mbox{}{r_+}^{8}+129024\,K \Bigl( {\frac {1381}{24}}+{b_{{2}}}^{3}+ \Bigl( {\frac {181}{18}}+b_{{3}} \Bigr) {b_{{2}}}^{2}
\nn\\
& \nonumber\mbox{}+ \Bigl( {\frac {73}{18}}\,b_{{3}}+{\frac {4825}{126}}+{\frac {4}{21}}\,{b_{{3}}}^{2}
\mbox{} \Bigr) b_{{2}} \Bigr) {T}^{4}{\pi }^{4}{r_+}^{7}-157696\,K{T}^{3} \Bigl( {\frac {2131}{77}}+{b_{{2}}}^{3}+ \Bigl({\frac {5679}{616}}+{\frac {54}{77}}\,b_{{3}} \Bigr) {b_{{2}}}^{2}
\nn\\
& \nonumber\mbox{}+ \Bigl({\frac {1299}{616}}\,b_{{3}}+{\frac {18087}{616}}+{\frac {3}{77}}\,{b_{{3}}}^{2}
\mbox{} \Bigr) b_{{2}} \Bigr) {\pi }^{3}{r_+}^{6}+258048\,K \Bigl(  \Bigl( MT\pi +{\frac {5}{32}} \Bigr) {b_{{2}}}^{3}+ \Bigl( {\frac {17}{21}}\,MT \Bigl( b_{{3}}+{\frac {562}{51}} \Bigr)\pi 
\nn\\
& \nonumber +{\frac {251}{224}}-{\frac {5}{224}}\,b_{{3}} \Bigr) {b_{{2}}}^{2}
\mbox{}+ \Bigl( \frac{2}{21}\,M \Bigl( {b_{{3}}}^{2}+{\frac {157}{6}}\,b_{{3}}+{\frac {6937}{24}} \Bigr) T\pi -{\frac {67}{224}}\,b_{{3}}
\nn\\
& \nonumber \mbox{}-\frac{1}{28}\,{b_{{3}}}^{2}+{\frac {601}{336}} \Bigr) b_{{2}}+\frac{3}{16}+{\frac {1793}{72}}\,MT\pi \Bigr){T}^{2}{\pi}^{2}{r_+}^{5}
\mbox{}-168960\,K T \Bigl(\Bigl( MT\pi -{\frac {27}{880}} \Bigr) {b_{{2}}}^{3}
\nn\\
& \nonumber +\Bigl( {\frac {17}{55}}\, \Bigl( b_{{3}}+{\frac {1609}{68}} \Bigr) 
\mbox{}MT\pi -{\frac {63}{880}}\,b_{{3}}-{\frac {63}{176}} \Bigr) {b_{{2}}}^{2}
\mbox{}+ \Bigl( -{\frac {4}{55}}\,M \Bigl( {b_{{3}}}^{2}+\frac{3}{16}\,b_{{3}}-201 \Bigr)T\pi 
\nn\\
&\nonumber-{\frac {63}{440}}\,b_{{3}}-{\frac {489}{440}} \Bigr) b_{{2}}-{\frac {57}{55}}+{\frac {961}{110}}\,MT\pi \mbox{} \Bigr) \pi \,{r_+}^{4}+153600\,K \Bigl(  \Bigl( {M}^{2}{T}^{2}{\pi }^{2}-{\frac {9}{1600}} \Bigr) {b_{{2}}}^{3}
\nn\\
&\nonumber\mbox{}+ \Bigl( -{\frac {27}{800}}+{\frac {13}{25}}\,{M}^{2}{T}^{2} \Bigl( b_{{3}}+{\frac {363}{26}} \Bigr) 
\mbox{}{\pi }^{2}-{\frac {21}{100}}\,MT \Bigl( b_{{3}}+{\frac {75}{28}} \Bigr)\pi  \Bigr) {b_{{2}}}^2 
\nn\\
&\nonumber+ \Bigl( -{\frac {27}{400}}+{\frac {39}{50}}\,{M}^{2}{T}^{2} \Bigl( b_{{3}}+{\frac {758}{39}} \Bigr) 
\mbox{}{\pi}^{2}-{\frac {147}{400}}\,MT \Bigl( b_{{3}}+{\frac {118}{21}} \Bigr) \pi  \Bigr) b_{{2}}
\nn\\
&\nonumber-{\frac {9}{200}}-{\frac {369}{200}}\,MT\pi 
\mbox{}+{\frac {489}{50}}\,{M}^{2}{T}^{2}{\pi }^{2} \Bigr) {r_+}^{3}
\nn\\
&\nonumber-27648\, \Bigl( b_{{2}}+\frac{3}{2} \Bigr) MK \Bigl(  \Bigl( MT\pi -\frac{1}{8} \Bigr) {b_{{2}}}^{2}+ \Bigl(-\frac{1}{2}-{\frac {7}{9}}\, \Bigl( b_{{3}}-{\frac {11}{7}} \Bigr)\mbox{}MT\pi  \Bigr) b_{{2}}
\nn\\
&\nonumber-\frac{1}{2}-{\frac {11}{18}}\,MT\pi \mbox{} \Bigr) {r_+}^{2}+24576\, \Bigl( b_{{2}}+\frac{3}{2} \Bigr) ^{2}{M}^{2}K \Bigl(  \Bigl( MT\pi -\frac{3}{16}\Bigr) b_{{2}}+\frac{3}{2}\,MT\pi -\frac{3}{8} \Bigr)r_+
\nn\\ 
&\nonumber+2048\, \Bigl( b_{{2}}+\frac{3}{2}\Bigr) ^{3}{M}^{3}K \Bigr]  \,.
\end{align}

The coefficients $d_1$ through $d_5$ appearing in the continued fraction approximation for $g(r)$ are as follows:

\be
d_1 = \frac{6M}{r_+} - g_0r_+^3, \quad d_2 = -\frac{24M+(A+3)r_+d_1+g_1r_+^5}{d_1r_+}
\ee

\be
d_{3}={\frac {2\,g_{2}\,{{r_+}}^{6}- \left( 2\,{d_{2}}^{2}+ \left( 
2\,A+8 \right) d_{2}+{A}^{2}+7\,A+12 \right) d_{1}\,{r_+}-20\,C}{2
\,d_{1}\,d_{2}\,{r_+}}}
\ee
\begin{align}
d_{4}=&\frac {1}{6\,d_{1
}\,d_{2}\,d_{3}\,{r_+}}\left[-6\,g_{3}\,{{r_+}}^{7}- \left( 6\,{d_{2}}^{3}+
 \left( 6\,A+12\,d_{3}+30 \right) {d_{2}}^{2}+ \left( 3\,{A}^{2}+
 \left( 6\,d_{3}+27 \right) A+6\,{d_{3}}^{2}+30\,d_{3}+60 \right) d_{2
} \right.\right. \\ \nonumber 
&\left.\left. +{A}^{3}+12\,{A}^{2}+47\,A+60 \right) d_{1}\,{r_+}-120\,C \right]
\end{align}

\begin{align}
d_{5}=&\frac {1}{24\,d_{1}\,d_{2}\,d_{3}\,d_
{4}\,{r_+}}\left[24\,g_{4}\,{{r_+}}^{8}- \left( 24\,{d_{2}}^{4}+
 \left( 24\,A+72\,d_{3}+144 \right) {d_{2}}^{3}+ \left( 72\,{d_{3}}^{2
}+ \left( 48\,A+48\,d_{4}+288 \right) d_{3}\right.\right.\right. \\ \nonumber 
&\left.\left.\left.+12\,{A}^{2}+132\,A+360
 \right) {d_{2}}^{2}+ \left( 24\,{d_{3}}^{3}+ \left( 24\,A+48\,d_{4}+
144 \right) {d_{3}}^{2}+ \left( 12\,{A}^{2}+ \left( 24\,d_{4}+132
 \right) A\right.\right.\right.\right. \\ \nonumber 
&\left.\left.\left.\left.+24\,{d_{4}}^{2}+144\,d_{4}+360 \right) d_{3}+4\,{A}^{3}+60
\,{A}^{2}+296\,A+480 \right) d_{2}+{A}^{4}+18\,{A}^{3}+119\,{A}^{2}+
342\,A\right.\right. \\ \nonumber 
&\left.\left.+360 \right) d_{1}\,{r_+}-840\,C\right]
\end{align}

\section{Appendix D: Solution properties for all theories}
\label{AppD}

Here we present perturbative expressions from the section on solution properties for the other five theories. Being derived from the field equations, the expressions are all equivalent for theories three and four.
\subsection{Photon Sphere}

Theory 1
\begin{align}
r_{ps (1)}&= 3M+{\frac {824\lambda_{(1)}}{32805{M}^{5}}}+{\frac {5959936{\lambda_{(1)}}^{2}}{3228504075{M}^{11}}}+a\left( -{\frac {2l_{z}}{9M}}+{\frac {5584l_{z}\lambda_{(1)}}{885735{M}^{7}}}+{\frac {15409408 l_{z}{\lambda_{(1)}}^{2}}{29056536675{M}^{13}}} \right) 
\\
 j_{ps (1)}^2&=27{M}^{2}+{\frac {104\lambda_{(1)}}{729{M}^{4}}}+{\frac {257792{\lambda_{(1)}}^{2}}{39858075{M}^{10}}}+a\left( -4l_{z}+{\frac {1136 l_{z}\lambda_{(1)}}{24057{M}^{6}}}+{\frac {2157479552 l_{z}{\lambda_{(1)}}^{2}}{224919117225{M}^{12}}} \right) 
\end{align}
\\
Theory 2     
\begin{align}
r_{ps (2)}&= 3M+{\frac {412\lambda_{(2)}}{6561{M}^{5}}}+{\frac {1489984 {\lambda_{(2)}}^{2}}{129140163 {M}^{11}}}+a\left( -{\frac {2l_{z}}{9M}}+{\frac { 4520 l_{z}\lambda_{(2)}}{ 177147 {M}^{7}}}+{\frac {278236832 l_{z}{\lambda_{(2)}}^{2}}{29056536675 {M}^{13}}} \right) 
\\
 j_{ps (2)}^2&=27{M}^{2}+{\frac {260 \lambda_{(2)}}{729{M}^{4}}}+{\frac { 64448 {\lambda_{(2)}}^{2}}{1594323 {M}^{10}}}+a\left( -4l_{z}+{\frac {3992 l_{z}\lambda_{(2)}}{24057{M}^{6}}}+{\frac {8322662752  l_{z}{\lambda_{(2)}}^{2}}{224919117225 {M}^{12}}} \right) 
\end{align}
\\

Theories 3, 4         
\begin{align}
r_{ps (4)}&= 3M+{\frac {1648\lambda_{(4)}}{32805{M}^{5}}}+{\frac {23839744 {\lambda_{(4)}}^{2}}{3228504075{M}^{11}}}+a\left( -{\frac {2l_{z}}{9M}}
+{\frac {11168 l_{z}\lambda_{(4)}}{885735{M}^{7}}} + {\frac {228442112 l_{z}{\lambda_{(4)}}^{2}}{29056536675  {M}^{13}}} \right) 
\\
 j_{ps (4)}^2&=27{M}^{2} +{\frac {208 \lambda_{(4)}}{729{M}^{4}}}+{\frac {1031168 {\lambda_{(4)}}^{2}}{39858075{M}^{10}}}+a\left( -4l_{z}+{\frac {2272  l_{z}\lambda_{(4)}}{24057{M}^{6}}} +{\frac {6646902272  l_{z}{\lambda_{(4)}}^{2}}{224919117225 {M}^{12}}} \right) 
\end{align}
  \\
Theory 5        
\begin{align}
r_{ps (5)}&= 3M+{\frac {3296\lambda_{(5)}}{32805{M}^{5}}}+{\frac {95358976 {\lambda_{(5)}}^{2}}{3228504075{M}^{11}}}+a\left( -{\frac {2l_{z}}{9M}}
+{\frac {39616 l_{z}\lambda_{(5)}}{885735{M}^{7}}} + {\frac {610285568  l_{z}{\lambda_{(5)}}^{2}}{29056536675  {M}^{13}}} \right) 
\\
 j_{ps (5)}^2&=27{M}^{2} +{\frac {416 \lambda_{(5)}}{729{M}^{4}}}+{\frac {4124672 {\lambda_{(5)}}^{2}}{39858075{M}^{10}}}+a\left( -4l_{z}+{\frac {6848  l_{z}\lambda_{(5)}}{24057{M}^{6}}} +{\frac {18529323008   l_{z}{\lambda_{(5)}}^{2}}{224919117225 {M}^{12}}} \right) 
\end{align}
   \\
Theory 6
\begin{align}
r_{ps (6)}&= 3M-{\frac {1648\lambda_{(6)}}{32805{M}^{5}}}+{\frac {23839744{\lambda_{(6)}}^{2}}{3228504075{M}^{11}}}+a\left( -{\frac {2l_{z}}{9M}}-{\frac {19808l_{z}\lambda_{(6)}}{885735{M}^{7}}}+{\frac {158446592l_{z}{\lambda_{(6)}}^{2}}{29056536675{M}^{13}}} \right) 
\\
 j_{ps (6)}^2&=27{M}^{2}-{\frac {208\lambda_{(6)}}{729{M}^{4}}}+{\frac {1031168{\lambda_{(6)}}^{2}}{39858075{M}^{10}}}+a\left( -4l_{z}-{\frac {3424l_{z}\lambda_{(6)}}{24057{M}^{6}}}+{\frac {4776791552l_{z}{\lambda_{(6)}}^{2}}{224919117225{M}^{12}}} \right) 
\end{align}

\subsection{ISCO parameters}

Theory 1     
\begin{align}
r_{\text{ISCO}(1)} &= 6M + {\frac {2221\lambda_{(1)}}{419904 {M}^{5}}}-{\frac {82796797{\lambda_{(1)}}^{2}}{5289581076480 {M}^{11}}}
\\ \nonumber &\mp  a\left(-{\frac {4\sqrt {6}}{3}}+ {\frac {8995\lambda_{(1)}\sqrt {6}}{1259712 {M}^{6}}}-{\frac {12794610641 {\lambda_{(1)}}^{2}\sqrt {6}}{317374864588800 {M}^{12}}}\right)
\\
j_{\text{ISCO}(1)}&= \pm\left(2\sqrt {3}M + \frac{1373\sqrt {3}\lambda_{(1)}}{6298560 {M}^{5}}-\frac {38293721\sqrt {3}\lambda_{(1)}^{2}}{79343716147200 {M}^{11}}\right)
\\ \nonumber &+ a\left(-{\frac {2\sqrt {2}}{3}} + {\frac {12901\sqrt {2}\lambda_{(1)}}{12597120 {M}^{6}}}-{\frac {2717557117 \sqrt {2}{\lambda_{(1)}}^{2}}{634749729177600 {M}^{12}}}\right)
\\
E_{\text{ISCO}(1)}&={\frac {2\sqrt {2}}{3}} + {\frac {191\sqrt {2}\lambda_{(1)}}{12597120 {M}^{6}}} - {\frac {26522027\sqrt {2}{\lambda_{(1)}}^{2}}{634749729177600{M}^{12}}}
\\ \nonumber 
&\mp a\left(-{\frac {\sqrt {3}}{54M}} + \frac{424843\sqrt {3}\lambda_{(1)}}{7482689280\,{M}^{7}} 
 - \frac{81838085603 \sqrt {3}{\lambda_{(1)}}^{2}}{298491060145766400 {M}^{13}}
 \right)
\end{align}
\\
Theory 2       
\begin{align} 
r_{\text{ISCO}(2)} &= 6M + {\frac {11105 \lambda_{(2)}}{839808 {M}^{5}}} - {\frac {413983985{\lambda_{(2)}}^{2}}{4231664861184 {M}^{11}}}
\\ \nonumber &\mp  a\left(-{\frac {4\sqrt {6}}{3}}+ {\frac {47855\lambda_{(2)}\sqrt {6}}{2519424 {M}^{6}}}-{\frac {74710165333 {\lambda_{(2)}}^{2}\sqrt {6}}{253899891671040 {M}^{12}}}\right)
\\
j_{\text{ISCO}(2)}&= \pm\left(2\sqrt {3}M + \frac{1373\sqrt {3}\lambda_{(2)}}{2519424 {M}^{5}}-\frac {38293721\sqrt {3}\lambda_{(2)}^{2}}{12694994583552 {M}^{11}}\right)
\\ \nonumber &+ a\left(-{\frac {2\sqrt {2}}{3}} + {\frac {13477 \sqrt {2}\lambda_{(2)}}{5038848  {M}^{6}}}-{\frac {75164507413 \sqrt {2}{\lambda_{(2)}}^{2}}{2538998916710400  {M}^{12}}}\right)
\\
E_{\text{ISCO}(2)}&={\frac {2\sqrt {2}}{3}} + {\frac {191\sqrt {2}\lambda_{(2)}}{5038848 {M}^{6}}} - {\frac {26522027\sqrt {2}{\lambda_{(2)}}^{2}}{101559956668416  {M}^{12}}}
\\ \nonumber 
&\mp a\left(-{\frac {\sqrt {3}}{54M}} + \frac{441259 \sqrt {3}\lambda_{(2)}}{2993075712 {M}^{7}} 
 - \frac{2231664953261  \sqrt {3}{\lambda_{(2)}}^{2}}{1193964240583065600 {M}^{13}}
 \right)
\end{align}
\\
Theories 3, 4      
\begin{align}
r_{\text{ISCO}(4)} &= 6M + {\frac {2221\lambda_{(4)}}{209952 {M}^{5}}}-{\frac {82796797{\lambda_{(4)}}^{2}}{1322395269120 {M}^{11}}}
\\ \nonumber &\mp  a\left(-{\frac {4\sqrt {6}}{3}}+ {\frac {8995\lambda_{(4)}\sqrt {6}}{629856 {M}^{6}}}-{\frac {13194848849 {\lambda_{(4)}}^{2}\sqrt {6}}{79343716147200 {M}^{12}}}\right)
\\
j_{\text{ISCO}(4)}&= \pm\left(2\sqrt {3}M + \frac{1373\sqrt {3}\lambda_{(4)}}{3149280 {M}^{5}}-\frac {38293721\sqrt {3}\lambda_{(4)}^{2}}{19835929036800 {M}^{11}}\right)
\\ \nonumber &+ a\left(-{\frac {2\sqrt {2}}{3}} + {\frac {12901\sqrt {2}\lambda_{(4)}}{6298560 {M}^{6}}}-{\frac {2763471229 \sqrt {2}{\lambda_{(4)}}^{2}}{158687432294400 {M}^{12}}}\right)
\\
E_{\text{ISCO}(4)}&={\frac {2\sqrt {2}}{3}} + {\frac {191\sqrt {2}\lambda_{(4)}}{6298560 {M}^{6}}} - {\frac {26522027\sqrt {2}{\lambda_{(4)}}^{2}}{158687432294400 {M}^{12}}}
\\ \nonumber 
&\mp a\left(-{\frac {\sqrt {3}}{54M}} + \frac{424843\sqrt {3}\lambda_{(4)}}{3741344640\,{M}^{7}} 
 - \frac{82940214983 \sqrt {3}{\lambda_{(4)}}^{2}}{74622765036441600  {M}^{13}}
 \right)
\end{align}
 \\
Theory 5 
\begin{align}
r_{\text{ISCO}(5)} &= 6M + {\frac {2221\lambda_{(5)}}{104976 {M}^{5}}}-{\frac {82796797{\lambda_{(5)}}^{2}}{330598817280 {M}^{11}}}
\\ \nonumber &\mp  a\left(-{\frac {4\sqrt {6}}{3}}- {\frac {96577\lambda_{(5)}\sqrt {6}}{1574640 {M}^{6}}}-{\frac {6673495717 {\lambda_{(5)}}^{2}\sqrt {6}}{3967185807360 {M}^{12}}}\right)
\\
j_{\text{ISCO}(5)}&= \pm\left(2\sqrt {3}M + \frac{1373\sqrt {3}\lambda_{(5)}}{1574640 {M}^{5}}-\frac {38293721\sqrt {3}\lambda_{(5)}^{2}}{4958982259200 {M}^{11}}\right)
\\ \nonumber &+ a\left(-{\frac {2\sqrt {2}}{3}} + {\frac {13621\sqrt {2}\lambda_{(5)}}{3149280 {M}^{6}}}-{\frac {3083362909 \sqrt {2}{\lambda_{(5)}}^{2}}{39671858073600 {M}^{12}}}\right)
\\
E_{\text{ISCO}(5)}&={\frac {2\sqrt {2}}{3}} + {\frac {191\sqrt {2}\lambda_{(5)}}{3149280 {M}^{6}}} - {\frac {26522027\sqrt {2}{\lambda_{(5)}}^{2}}{39671858073600 {M}^{12}}}
\\ \nonumber 
&\mp a\left(-{\frac {\sqrt {3}}{54M}} + \frac{445363\sqrt {3}\lambda_{(5)}}{1870672320\,{M}^{7}} 
 - \frac{91231491983 \sqrt {3}{\lambda_{(5)}}^{2}}{18655691259110400  {M}^{13}}
 \right)
\end{align}
   \\
Theory 6 
\begin{align}
r_{\text{ISCO}(6)} &= 6M-{\frac {2221\lambda_{(6)}}{209952{M}^{5}}}-{\frac {82796797{\lambda_{(6)}}^{2}}{1322395269120{M}^{11}}}
\\ \nonumber &\mp  a\left(-{\frac {4\sqrt {6}}{3}}-{\frac {9715\lambda_{(6)}\sqrt {6}}{629856{M}^{6}}}-{\frac {15486310769{\lambda_{(6)}}^{2}\sqrt {6}}{79343716147200{M}^{12}}}\right)
\\
j_{\text{ISCO}(6)}&= \pm\left(2\sqrt {3}M-\frac{1373\sqrt {3}\lambda_{(6)}}{3149280{M}^{5}}-\frac {38293721\sqrt {3}\lambda_{(6)}^{2}}{19835929036800{M}^{11}}\right)
\\ \nonumber &+ a\left(-{\frac {2\sqrt {2}}{3}}-{\frac {13621\sqrt {2}\lambda_{(6)}}{6298560{M}^{6}}}-{\frac {3080166109\sqrt {2}{\lambda_{(6)}}^{2}}{158687432294400{M}^{12}}}\right)
\\
E_{\text{ISCO}(6)}&={\frac {2\sqrt {2}}{3}}-{\frac {191\sqrt {2}\lambda_{(6)}}{6298560{M}^{6}}}-{\frac {26522027\sqrt {2}{\lambda_{(6)}}^{2}}{158687432294400{M}^{12}}}
\\ \nonumber &\mp a\left(-{\frac {\sqrt {3}}{54M}}-{\frac {445363\sqrt {3}\lambda_{(6)}}{3741344640\,{M}^{7}}}-{\frac {91154747183\sqrt {3}{\lambda_{(6)}}^{2}}{74622765036441600{M}^{13}}}
\right)
\end{align}

\subsection{Photon ring parameters}

Theory 1
\begin{align}
r_{\text{pr}\pm (1) } &= 3M + {\frac {824\lambda_{(1)}}{32805{M}^{5}}} + {\frac {5959936{\lambda_{(1)}}^{2}}{3228504075{M}^{11}}}\\ \nonumber &\pm a\left( {\frac {2\sqrt {3}}{3}} - {\frac {1688\sqrt {3}\lambda_{(1)}}{98415{M}^{6}}}-{\frac {76780144 \sqrt {3}{\lambda_{(1)}}^{2}}{9685512225{M}^{12}}} \right) 
\\
\omega_{\text{pr}\pm (1)}&=\pm\left(-{\frac {\sqrt {3}}{9\,M}}+{\frac {52\,\sqrt {3}\lambda_{(1)}}{177147\,{M}^{7}}}+{\frac {352888\,\sqrt {3}{\lambda_{(1)}}^{2}}{29056536675\,{M}^{13}}}\right)\\ \nonumber & + a\left( {\frac {2}{27\,{M}^{2}}} - {\frac {7400\,\lambda_{(1)}}{5845851\,{M}^{8}}}-{\frac {937537408\,{\lambda_{(1)}}^{2}}{4968667771425 \,{M}^{14}}} \right) 
\end{align}
\\
Theory 2
\begin{align}
r_{\text{pr}\pm (2) } &= 3M + {\frac {412 \lambda_{(2)}}{6561{M}^{5}}} + {\frac {1489984 {\lambda_{(2)}}^{2}}{129140163 {M}^{11}}}\\ \nonumber &\pm a\left( {\frac {2\sqrt {3}}{3}} - {\frac {1420 \sqrt {3}\lambda_{(2)}}{19683 {M}^{6}}}-{\frac {278440732 \sqrt {3}{\lambda_{(2)}}^{2}}{9685512225{M}^{12}}} \right) 
\\
\omega_{\text{pr}\pm (2)}&=\pm\left(-{\frac {\sqrt {3}}{9\,M}}+{\frac {130\,\sqrt {3}\lambda_{(2)}}{177147\,{M}^{7}}}+{\frac {88222\,\sqrt {3}{\lambda_{(2)}}^{2}}{1162261467 \,{M}^{13}}}\right)\\ \nonumber & + a\left( {\frac {2}{27\,{M}^{2}}} - {\frac {23684\,\lambda_{(2)}}{5845851\,{M}^{8}}}-{\frac {3689849344\,{\lambda_{(2)}}^{2}}{4968667771425 \,{M}^{14}}} \right) 
\end{align}
\\
Theories 3,4
\begin{align}
r_{\text{pr}\pm (4) } &= 3M + {\frac {1648\lambda_{(4)}}{32805{M}^{5}}}+{\frac {23839744{\lambda_{(4)}}^{2}}{3228504075{M}^{11}}}\\ \nonumber &\pm a\left( {\frac {2\sqrt {3}}{3}} - {\frac {3376\sqrt {3}\lambda_{(4)}}{98415{M}^{6}}}-{\frac {102663616\sqrt {3}{\lambda_{(4)}}^{2}}{9685512225{M}^{12}}} \right) 
\\
\omega_{\text{pr}\pm (4)}&=\pm\left(-{\frac {\sqrt {3}}{9\,M}} + {\frac {104\,\sqrt {3}\lambda_{(4)}}{177147\,{M}^{7}}}+{\frac {1411552\,\sqrt {3}{\lambda_{(4)}}^{2}}{29056536675\,{M}^{13}}}\right)\\ \nonumber & + a\left( {\frac {2}{27\,{M}^{2}}}- {\frac {14800\,\lambda_{(4)}}{5845851\,{M}^{8}}}-{\frac {1693553152\,{\lambda_{(4)}}^{2}}{4968667771425 \,{M}^{14}}} \right) 
\end{align}
\\
Theory 5
\begin{align}
r_{\text{pr}\pm (5) } &= 3M + {\frac {3296\lambda_{(5)}}{32805{M}^{5}}}+{\frac {95358976{\lambda_{(5)}}^{2}}{3228504075{M}^{11}}}\\ \nonumber &\pm a\left( {\frac {2\sqrt {3}}{3}}-{\frac {12512\sqrt {3}\lambda_{(5)}}{98415{M}^{6}}} - {\frac {612005632\sqrt {3}{\lambda_{(5)}}^{2}}{9685512225{M}^{12}}} \right) 
\\
\omega_{\text{pr}\pm (5)}&=\pm\left(-{\frac {\sqrt {3}}{9\,M}} + {\frac {208\,\sqrt {3}\lambda_{(5)}}{177147\,{M}^{7}}}+{\frac {5646208\,\sqrt {3}{\lambda_{(5)}}^{2}}{29056536675\,{M}^{13}}}\right)\\ \nonumber & + a\left( {\frac {2}{27\,{M}^{2}}} - {\frac {39968\,\lambda_{(5)}}{5845851\,{M}^{8}}}-{\frac {330913792\,{\lambda_{(5)}}^{2}}{198746710857\,{M}^{14}}} \right) 
\end{align}
\\
Theory 6
\begin{align}
r_{\text{pr}\pm (6) } &= 3M-{\frac {1648\lambda_{(6)}}{32805{M}^{5}}}+{\frac {23839744{\lambda_{(6)}}^{2}}{3228504075{M}^{11}}}\\ \nonumber &\pm a\left( {\frac {2\sqrt {3}}{3}}+{\frac {6256\sqrt {3}\lambda_{(6)}}{98415{M}^{6}}}-{\frac {158876608\sqrt {3}{\lambda_{(6)}}^{2}}{9685512225{M}^{12}}} \right) 
\\
\omega_{\text{pr}\pm (6)}&=\pm\left(-{\frac {\sqrt {3}}{9\,M}}-{\frac {104\,\sqrt {3}\lambda_{(6)}}{177147\,{M}^{7}}}+{\frac {1411552\,\sqrt {3}{\lambda_{(6)}}^{2}}{29056536675\,{M}^{13}}}\right)\\ \nonumber & + a\left( {\frac {2}{27\,{M}^{2}}}+{\frac {19984\,\lambda_{(6)}}{5845851\,{M}^{8}}}-{\frac {85092352\,{\lambda_{(6)}}^{2}}{198746710857\,{M}^{14}}} \right) 
\end{align}

\subsection{Black Hole Shadow parameters}
 
Theory 1
\begin{align}
R_{\text{sh} (1)}&=3\sqrt {3}M + {\frac {52\sqrt {3}}{6561{M}^{5}}}\lambda_{(1)}+{\frac {1126264\sqrt {3}}{3228504075{M}^{11}}}{\lambda_{(1)}}^{2} 
\\
D_{(1)} &= a\sin(\theta_0)\left(-2 + {\frac {568}{24057{M}^{6}}}\lambda_{(1)}+{\frac {1078739776}{224919117225{M}^{12}}}{\lambda_{(1)}}^{2}\right)
\end{align} 
\\
Theory 2
\begin{align}
R_{\text{sh} (2)}&=3\sqrt {3}M + {\frac {130\sqrt {3}}{6561{M}^{5}}}\lambda_{(2)}+{\frac {281566\sqrt {3}}{129140163 {M}^{11}}}{\lambda_{(2)}}^{2} 
\\
D_{(2)} &= a\sin(\theta_0)\left(-2 + {\frac {1996}{24057{M}^{6}}}\lambda_{(2)}+{\frac {4161331376}{224919117225{M}^{12}}}{\lambda_{(2)}}^{2}\right)
\end{align} 
\\
Theories 3,4
\begin{align}
R_{\text{sh} (4)}&=3\sqrt {3}M + {\frac {104\sqrt {3}}{6561{M}^{5}}}\lambda_{(4)}+{\frac {4505056\sqrt {3}}{3228504075{M}^{11}}}{\lambda_{(4)}}^{2} 
\\
D_{(4)} &= a\sin(\theta_0)\left(-2+{\frac {1136}{24057{M}^{6}}}\lambda_{(4)}+{\frac {3323451136}{224919117225{M}^{12}}}{\lambda_{(4)}}^{2}\right)
\end{align}
\\ 
 Theory 5 
\begin{align}
R_{\text{sh} (5)}&=3\sqrt {3}M + {\frac {208\sqrt {3}}{6561{M}^{5}}}\lambda_{(5)}+{\frac {18020224\sqrt {3}}{3228504075{M}^{11}}}{\lambda_{(5)}}^{2} 
\\
D_{(5)} &= a\sin(\theta_0)\left(-2 + {\frac {3424}{24057{M}^{6}}}\lambda_{(5)}+{\frac {9264661504}{224919117225{M}^{12}}}{\lambda_{(5)}}^{2}\right)
\end{align}
\\ 
 Theory 6 
\begin{align}
R_{\text{sh} (6)}&=3\sqrt {3}M-{\frac {104\sqrt {3}}{6561{M}^{5}}}\lambda_{(6)}+{\frac {4505056\sqrt {3}}{3228504075{M}^{11}}}{\lambda_{(6)}}^{2} 
\\
D_{(6)} &= a\sin(\theta_0)\left(-2-{\frac {1712}{24057{M}^{6}}}\lambda_{(6)}+{\frac {2388395776}{224919117225{M}^{12}}}{\lambda_{(6)}}^{2}\right)
\end{align}

\bibliographystyle{ieeetr}
\bibliography{LBIB}

\end{document}